\documentclass[aps,footinbib,prl,superscriptaddress,reprint]{revtex4-1}
\usepackage{graphicx}
\usepackage{amssymb}
\usepackage{hyperref}
\usepackage{amsmath}
\usepackage{slashed}
\usepackage{lipsum}
\usepackage[dvipsnames]{xcolor}
\usepackage{xcolor}   
\usepackage{xspace} 
\usepackage{bm}
\usepackage{xfrac}
\usepackage{lineno}
\usepackage[normalem]{ulem}
\usepackage[toc,page]{appendix}
\usepackage[capitalize]{cleveref}
\usepackage{nicefrac}
\usepackage{relsize}

\usepackage[T1]{fontenc}      
\usepackage{ae}

\usepackage{grffile}

\RequirePackage{fix-cm} 
\DeclareMathSizes{9.5}{9.5}{7.0}{5}

\begin{document}


\title{Functional dependence of the Hall viscosity-induced transverse voltage in two-dimensional Fermi liquids}


\author{Ioannis~Matthaiakakis}
\thanks{These three authors contributed equally to this work and are listed alphabetically.}
\affiliation{Institute for Theoretical Physics and Astrophysics and W\"urzburg-Dresden Cluster of Excellence ct.qmat,
Julius-Maximilians-Universit\"at W\"urzburg, 97074 W\"urzburg, Germany}

\author{David~Rodr\'iguez~Fern\'andez}
\thanks{These three authors contributed equally to this work and are listed alphabetically.}
\affiliation{Institute for Theoretical Physics and Astrophysics and W\"urzburg-Dresden Cluster of Excellence ct.qmat,
Julius-Maximilians-Universit\"at W\"urzburg, 97074 W\"urzburg, Germany}

\author{\mbox{Christian~Tutschku}}
\thanks{These three authors contributed equally to this work and are listed alphabetically.}
\affiliation{Institute for Theoretical Physics and Astrophysics and W\"urzburg-Dresden Cluster of Excellence ct.qmat,
Julius-Maximilians-Universit\"at W\"urzburg, 97074 W\"urzburg, Germany}

\author{Ewelina~M. Hankiewicz}
\affiliation{Institute for Theoretical Physics and Astrophysics and W\"urzburg-Dresden Cluster of Excellence ct.qmat,
Julius-Maximilians-Universit\"at W\"urzburg, 97074 W\"urzburg, Germany}

\author{Johanna~Erdmenger}
\affiliation{Institute for Theoretical Physics and Astrophysics and W\"urzburg-Dresden Cluster of Excellence ct.qmat,
Julius-Maximilians-Universit\"at W\"urzburg, 97074 W\"urzburg, Germany}

\author{Ren\'e~Meyer}
\affiliation{Institute for Theoretical Physics and Astrophysics and W\"urzburg-Dresden Cluster of Excellence ct.qmat,
Julius-Maximilians-Universit\"at W\"urzburg, 97074 W\"urzburg, Germany}

\begin{abstract}
\noindent
The breaking of parity and time-reversal symmetry in two-dimensional Fermi liquids gives rise to non-dissipative transport features characterized by the Hall viscosity. In  magnetic fields, the Hall viscous force directly competes with the Lorentz force, since both mechanisms contribute to the Hall voltage. In this work, we present a channel geometry that allows us to uniquely distinguish these two contributions and derive, for the first time, their functional dependency on all external parameters. 
We show that the ratio of the Hall viscous to the Lorentz force contribution is negative and that its modulus decreases with increasing width, slip-length and carrier density, while it increases with the electron-electron mean free path of our channel. 
In typical materials such as GaAs the Hall viscous contribution can dominate the Lorentz signal up to a few tens of millitesla until the total Hall voltage vanishes and subsequently is overcome by the Lorentz contribution. Moreover, we prove that  the  total Hall electric field is parabolic due to Lorentz effects, whereas the offset of this parabola is  characterized by the Hall viscosity. 
Hence, our results pave the way to measure and identify the Hall viscosity via both global and local voltage measurements.
\end{abstract}


\maketitle


\textit{Introduction.}
The idea of describing electrons in solids via  hydrodynamics goes back to the discovery of the Gurzhi effect in (Al)GaAs quantum wires
\cite{Molenkamp:1994ii,Molenkamp:1994kb,deJong:1995bn}. Recently, hydrodynamic
transport has received renewed attention due to the accessibility of the hydrodynamic regime in modern materials \cite{Lucas2018,bandurin2016negative,KrishnaKumar2017,Moll:2016ju}, even beyond the Fermi liquid regime \cite{Gusev18,Gusev18:2,Bakarov18}. In particular, two-dimensional systems that violate parity invariance 
are of special interest, since they exhibit novel non-dissipative hydrodynamic transport coefficients, such as the Hall viscosity $\eta_\mathrm{H}$ \cite{AvronSeilerZograf,Hoyos:2014pba,alekseev2016negative,Schmalian17,Surowka19,Landsteiner19}. Recent experiments in graphene have shown that $\eta_{\rm H}$ may be of the same order of magnitude as the shear viscosity $\eta$ and therefore is expected to significantly affect the fluid transport \cite{Berdyugineaau0685}.
However, current theoretical and experimental works
\cite{Delacretaz:2017yia,Pellegrino:2017fd,Stern2019,2019arXiv190511662S} do not provide a quantitative answer to the functional dependency of Hall viscous \mbox{effects} on all parameters describing the system, including a finite slip- $l_\mathrm{s}$ and impurity length $l_\mathrm{imp}$.

\begin{figure}[t]
\centering
\includegraphics[scale=0.235]{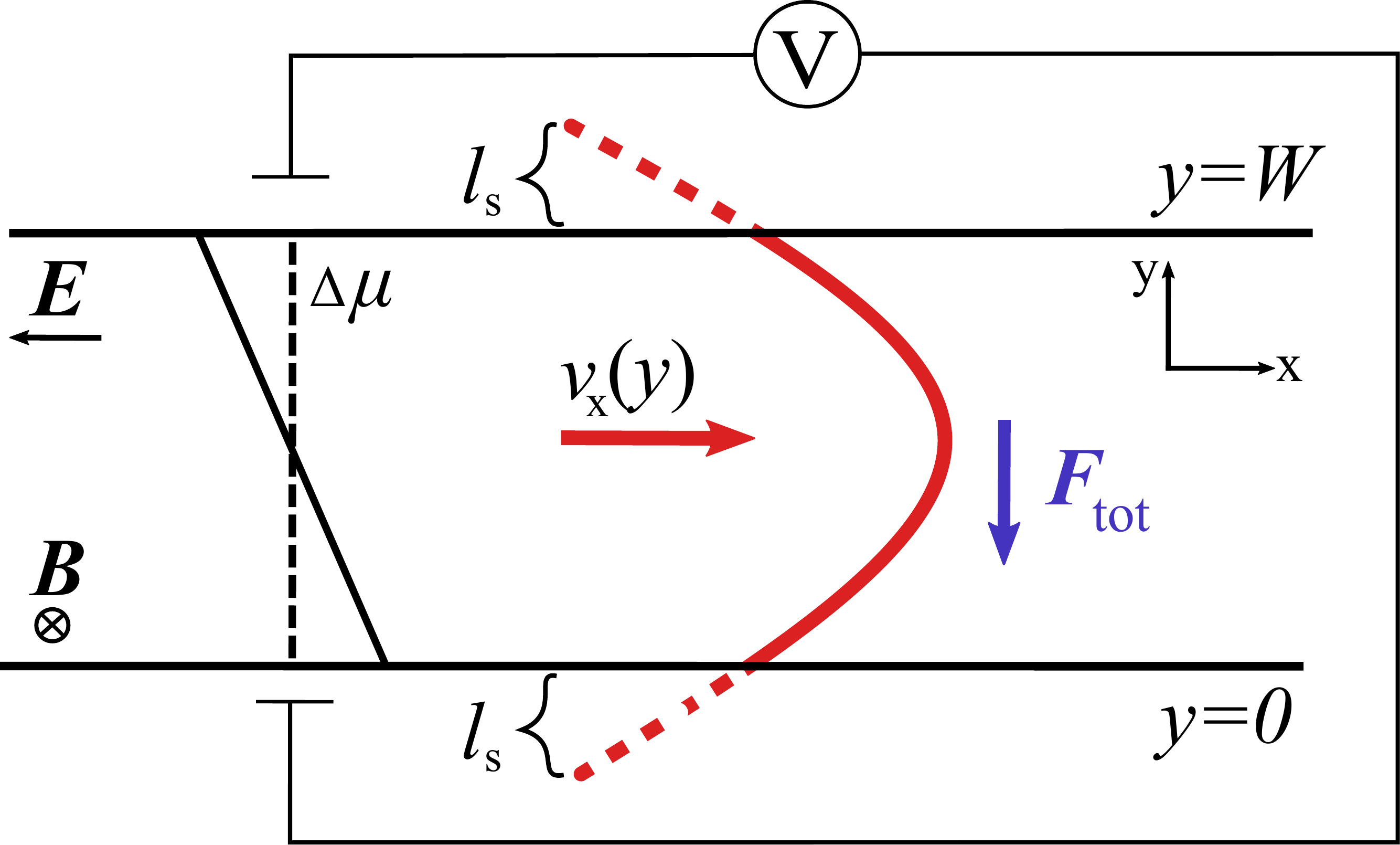}
\caption{Setup to distinguish Hall viscous from Lorentz force contributions to the total Hall voltage. The red curve shows the hydrodynamic velocity profile $v_\mathrm{x}(y)$ in a channel of length $L$ and width $W \! \! \ll \! \! L$ in presence of an electric field $\textbf{E}$  and an out-of-plane magnetic field $\mathbf{B}$. Momentum transfer to the boundaries, captured by the slip length $l_\mathrm{s}$, is shown by the red dashed curve. The total Hall force $\textbf{F}_\mathrm{_{\! tot}}$ induces a transverse pressure gradient, giving rise to a gradient in the chemical potential $\Delta\mu$, illustrated (initially) by the (dashed)  black line.}
\label{fig:poisetah}
\vspace{-.0cm}
\end{figure}

To address this open issue, we predict in this work  the Hall viscous contribution $\Delta  V_{\eta_\mathrm{H}}$ to the total  Hall voltage $\Delta  V_\mathrm{tot} = \Delta
V_{\eta_\mathrm{H}} + \Delta  V_{B}$ measured across a
two-dimensional channel in the hydrodynamic regime. Here,  $\Delta  V_{B}$ is the Lorentz force contribution. 
Up to first order in the electric field, we derive the complete functional dependency of $\Delta  V_{\eta_\mathrm{H}}$ and $\Delta  V_B$ on all external parameters  as a function of the transverse channel coordinate.
In~particular, we evaluate the dependence of both voltage contributions on $l_\mathrm{imp}$, $l_\mathrm{s}$, 
the wire width $W$, the equilibrium carrier density $\rho_0$ as well as on the magnetic field $B$.
This allows to distinguish $\Delta  V_{\eta_{\rm H}}$  from  $\Delta V_{B}$ via measurements on two-dimensional Hall setups  with varying $W$ and $\rho_0$, such as shown in Fig.~\ref{fig:poisetah}. Most remarkably, we derive that the two voltage contributions differ in sign and that  the ratio $\vert \Delta  V_{\eta_{\rm H}} \vert / \vert \Delta V_{B} \vert $ decreases with increasing $\vert B \vert$. Hence, the total Hall voltage cancels at a certain critical magnetic field $B_\mathrm{c}$, which is a smoking gun feature of \mbox{Hall viscous transport.}
Additionally, we show that local measurements of the Hall voltage are suitable to identify the Hall viscosity. In particular, we analytically derive the parabolic form in the transverse channel coordinate of the total Hall electric field, as was recently measured in Ref.~\cite{2019arXiv190511662S}.
While the curvature of this parabola is mainly defined by the Lorentz contribution, its offset is explicitly characterized by $ \Delta V_{\eta_\mathrm{H}}$ and hence by the Hall viscosity.

For both, weak and strong magnetic fields as well as for clean systems our approach is analytical, whereas for intermediate field regimes we numerically solve the Navier-Stokes equations.
In particular, we demonstrate that $\Delta  V_B$ is proportional to the velocity profile integrated over the channel width, and therefore proportional to the total fluid momentum. In contrast, $\Delta V_{\eta_\mathrm{H}}$ exclusively depends on the gradient of the velocity profile at the boundaries of the system. Based on this, we analytically prove that for small fields and clean systems with Poiseuille-like (parabolic) velocity profile the ratio
$\Delta V_{\eta_\mathrm{H}}/\Delta  V_{B}$ is determined completely by the interplay of length scales defining the system. While the modulus of this ratio increases with the transverse electron-electron mean free path $l_\mathrm{ee}$, it decreases with $W,l_\mathrm{s}$ and $l_\mathrm{imp}$. 
Since $l_\mathrm{ee}$ and $l_\mathrm{imp}$ are density dependent, the ratio acquires a density dependence, as well. In the absence of impurities $\Delta V_{\eta_{\rm H}}/ \Delta V_{B} \! = \! \mathcal{O}(\rho_0^3)$, whereas for weak impurity strength there exists an additional $\mathcal{O}(\rho_0^2)$ contribution. 
Hence, it is possible to achieve $ \vert \Delta V_{\eta_\mathrm{H}} \vert / \vert \Delta  V_{B} \vert  \! \gg \! 1$ by tuning the width and density of the sample. For $\rho_0 \! =  \! 9.1  \times  10^{11} \mathrm{cm}^{-1}$ and $W  \! = \!  3 \mu$m, such a regime is for instance realized in GaAs for $ \vert B \vert  \! \leq \!  20$mT.
Beyond the weak field limit, we  show that increasing $\vert B \vert$ strongly reduces $ \vert \Delta V_{\eta_\mathrm{H}} \vert / \vert \Delta  V_{B} \vert$,  due to the  suppression of $\eta$ and $\eta_\mathrm{H}$  \cite{alekseev2016negative}.  
For large magnetic fields, for example in GaAs for $ \vert B \vert \! \gtrsim \! 0.5$T, impurities dominate the transport, which causes an Ohmic (flat) fluid profile with vanishing $ \Delta V_{\eta_\mathrm{H}}/\Delta  V_{B}$. The critical field $ \vert B_\mathrm{c} \vert$, at which  $ \Delta V_{\eta_\mathrm{H}}/ \Delta  V_{B} \! = \! -1$ and the total Hall voltage $\Delta V_\mathrm{tot}=0$, increases with decreasing $W,l_\mathrm{s},l_\mathrm{imp}$ and $l_\mathrm{ee}^{-1}$. In particular, for GaAs we find $ \vert B_\mathrm{c} \vert  \! \simeq \! \mathcal{O}(10 \mathrm{mT})$, which makes our predictions experimentally verifiable.

\vspace{.2cm}

\emph{Model.} Our setup is shown in  Fig.~\ref{fig:poisetah}. 
We analyze the hydrodynamic flow of non-relativistic electrons along a two-dimensional channel under the combined influence of a DC electric field $\mathbf{E} = -E_\mathrm{x} {\bf e}_\mathrm{x}$ and an out-of-plane magnetic field $\mathbf{B} = -B {\bf e}_\mathrm{z}$   \footnote{In what follows, we assume that the screening length in our material is much shorter than the width of our channel. Hence, the electric field sourced by the redistribution of charge carriers in our fluid can be neglected.}.
To justify our hydrodynamic approach, we assume that $l_\mathrm{ee}$ is the shortest length scale in our system \cite{erdmenger2018strongly}.   
In particular, it is smaller than the cyclotron radius $r_\mathrm{c} \! = \! m_\mathrm{eff} v_\mathrm{F}/\vert \mathrm{e} B  \vert $, which is defined in terms of the effective electron mass $m_\mathrm{eff}$, the electron charge $\mathrm{e}<0$ and the Fermi velocity $v_\mathrm{F}$.	Additionally, we assume that our system is incompressible and has translational invariance in longitudinal, as well as vanishing current in transversal direction. 
These assumptions lead to the hydrodynamic variables
\footnote{The charge conservation equation $\partial_t \rho + \nabla\cdot(\rho \mathbf{v})=0$ is trivially satisfied due to our assumption of a steady and translational invariant flow $\mathbf{v}=v_\mathrm{x}(y) \mathbf{e}_\mathrm{x}$.}
\begin{align}
\mathbf{v} = v_{\rm x}(y) \, \mathbf{e_{\rm x}},  \ \ \mu \! = \! \mu_0 \! + \! \Delta \mu, \ \ T = T_0 = \mathrm{constant} \ ,
\end{align}
where $\Delta \mu$ is the local chemical potential fluctuation around the global equilibrium value $\mu_0$ \footnote{In general, one could also allow for a temperature gradient $T \! \to \!  T \! + \!  \Delta T$. 
However, we show in Ref.~\cite{supp} that this contribution is negligibly small for Fermi liquids.
In addition, since we restrict ourselves to $T = \mathcal{O}(1{\rm K})$, phonons are negligible in our approach  \cite{Molenkamp:1994ii,Molenkamp:1994kb}.}.
For this ansatz, the dynamical equations are defined by the non-relativistic Navier-Stokes equations for incompressible fluids \cite{Landau:1987,Romatschke:2010jf,Rezzolla:2013,supp}. 
In the framework of linear response theory with respect to the external electric field $E_\mathrm{x}$, these equations are given by \footnote{Let us emphasize that we took into account the term $\propto \! \Delta \mu E_\mathrm{x}$, even though it is higher order in $E_\mathrm{x}$ and therefore does not affect our linear response results. We did so to explicitly see that the Lorentz and the Hall-viscosity induced force influences the velocity profile in the expected way (cf. Ref.~[29]).}
\begin{align}
\eta \, \partial^2_\mathrm{y} \, v_\mathrm{x} & = \mathrm{e} \left( \rho_0 +  \frac{\partial \rho_0}{\partial\mu}  \Delta  \mu \right)E_\mathrm{x} +\frac{\rho_0 v_\mathrm{F} m_\mathrm{eff}}{l_\mathrm{imp} }v_\mathrm{x},\, 
  \label{eq:betaxeq}\\
\partial_\mathrm{y} \, p \, & =    \rho_0 \, \partial_\mathrm{y} \Delta \mu = \left( \, {\rm e}B \rho_0  -\eta_\mathrm{H} \partial^2_\mathrm{y} \, \right) v_\mathrm{x}\, . \label{eq:chemgradeq}
\end{align}
Here, \mbox{$p \! = \!  p_0 \! + \! \rho_0 \Delta \mu$} defines the pressure in terms of $\Delta \mu$. 
Let us emphasize that the incompressibility condition for our unidirectional fluid flow, $\nabla \cdot \mathbf{v} \! = \! 0$,  solely implies a constant density along the $\mathbf{e}_\mathrm{x}$-direction. In particular, $v_\mathrm{y}=0$ allows for density fluctuations satisfying $\partial_\mathrm{y} \rho \propto \partial_\mathrm{y} \Delta \mu \neq 0$. 
In two dimensions, the dynamics of incompressible, non-relativistic fluids is entirely captured by the shear and Hall viscosities \cite{scaffidi2017hydrodynamic,alekseev2016negative}
\begin{align}
\label{eq:eta}
\eta = \frac{\eta_0}{1+(2 \, l_\mathrm{ee}/r_\mathrm{c})^2} \,, \quad \eta_\mathrm{H} = \frac{2 \, \mathrm{sgn}(B) \, \eta_0 \, l_\mathrm{ee}/r_\mathrm{c} }{1+(2 \, l_\mathrm{ee}/r_\mathrm{c})^2} \, , 
\end{align}
where $\eta_0 \! = \! m_\mathrm{eff} \rho_0   v_\mathrm{F}   l_{\rm ee}/4$ is the shear viscosity at  zero magnetic field. The main goal of this work is to derive the total Hall voltage
\begin{align} \label{defofVtot}
\Delta V_{\rm tot} & = - \left( \Delta \mu(W)-\Delta \mu(0) \right) / \mathrm{e}
\end{align}
up to first order in $E_\mathrm{x}$, and to separate Hall viscous and Lorentz force contributions therein. Therefore, we solve Eqs.~\eqref{eq:betaxeq} and~\eqref{eq:chemgradeq} under the general boundary conditions 
\begin{align} \label{eq:bsc}
v_{\rm x}(-l_{\rm s}) =  v_{\rm x}(W \! + \! l_{\rm s}) & =  0     , \quad  \Delta \mu \big\vert_{y=0}    = - \Delta \mu\big\vert_{y=W} \ .   
\end{align}
The slip length $l_\mathrm{s}$ characterizes the velocity profile at the boundaries, as shown in Fig.~\ref{fig:poisetah}. For $l_\mathrm{s}\!=\!0$, the fluid flow is Poiseuille-like (parabolic), while $l_\mathrm{s} \! \to \! \infty$ defines the diffusive regime with a flat $v_\mathrm{x}(y)$ \footnote{ Notice that we do not use standard Robin type boundary conditions to define the slip length in our system \cite{Kiselev18}. This relies on the fact that our set of boundary conditions naturally describes the case of a finite drift velocity.
In our supplemental material we  prove that in our channel geometry, Robin-type boundary conditions are equivalent with our choice of boundary conditions, Eq.\eqref{eq:bsc}.}.

\vspace{.2cm}

\textit{Hall Response.} 
The total Hall voltage $\Delta   V_\mathrm{tot} \! = \!  \Delta    V_{\eta_\mathrm{H}} \!+\Delta  V_{B}$ consists of two different terms, namely the Hall viscous and the Lorentz force contributions.
To analytically derive the functional dependence of these building blocks, we first restrict ourselves to weak  magnetic fields, defined by $l_\mathrm{G}/r_\mathrm{c} \! \ll \! 1$. The Gurzhi length \mbox{$l_{\rm G} \! =  \! \sqrt{ l_{\rm imp} \eta / (\rho_0 m_{\rm eff} v_{\rm F}  ) }$} quantifies the relative strength of impurity to shear effects. While the flow is Poiseuille-like for $l_{\rm G}/W \! \gg \! 1$, it becomes Ohmic for $l_{\rm G}/W \! \ll \! 1 $. 
The assumption of weak magnetic fields allows us to expand the velocity profile and hence the Navier-Stokes equations in powers of $B$. 
\begin{figure}[t!]
\centering
\vspace{.1cm}
	\includegraphics[scale=0.16]{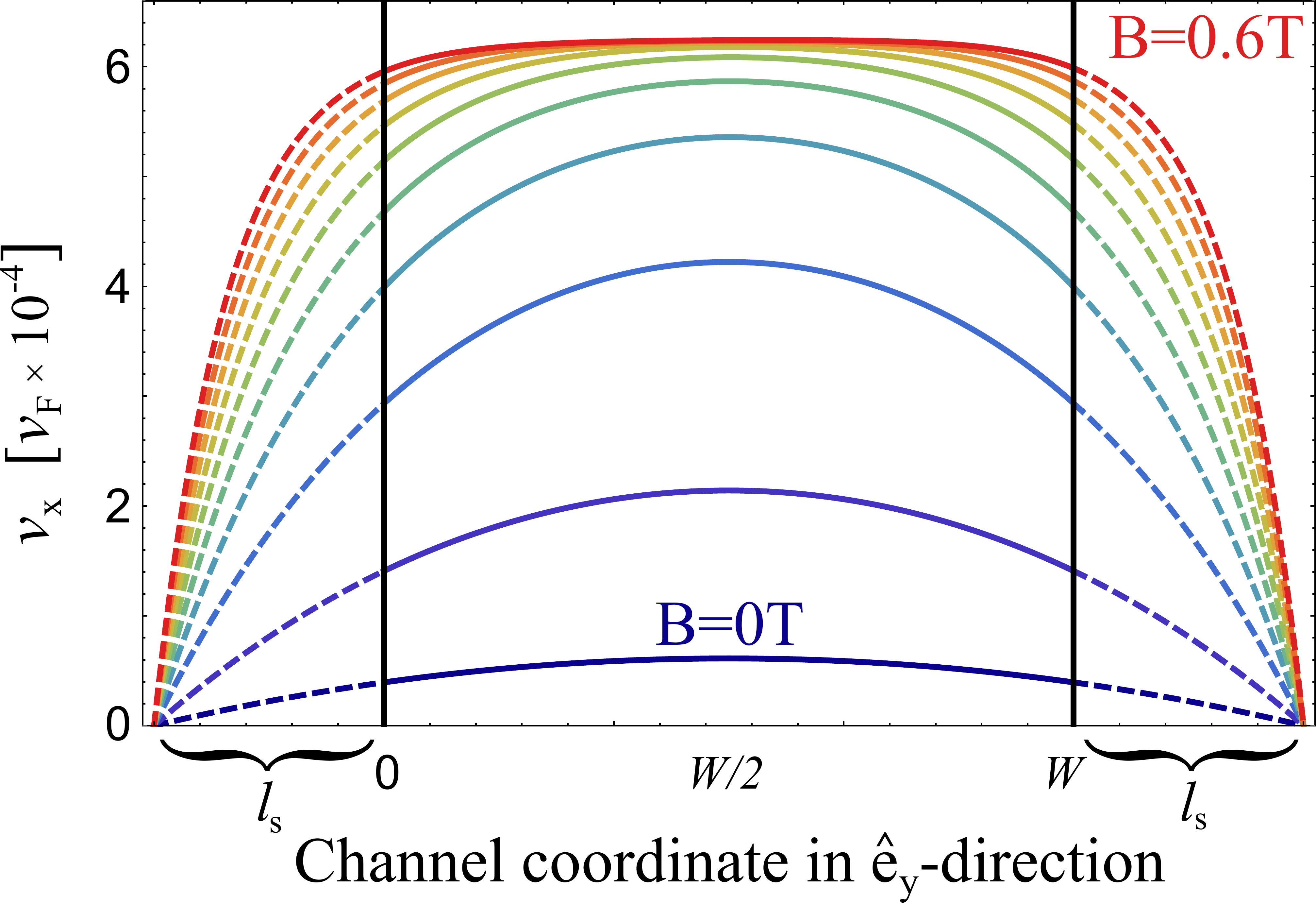}
	\caption{Velocity profile $v_\mathrm{x}(y)$ of a general Fermi liquid with $\rho_0\!=\!9.1\times 10^{11} \rm cm^{-2}$,  $\eta_0 \! = \! 1.7 \!\times \! 10^{-16} \rm{J s/m^2}$ and $l_{\rm imp} \! = \! 40 \mu \rm m$ \cite{Gusev18}.
From the bottom up, the magnetic field is raised from $B=0 $T (blue) to $B\! = \!0.6$T (red), associated with an  increase of the fluid velocity. The asymptotic  flat velocity profile is caused by impurities, since for large magnetic fields these provide the only mechanism for attaining a steady fluid flow. }
\label{fig:poiseuille}
\end{figure}
With this ansatz, the linearized Navier-Stokes \mbox{Eqs.~\eqref{eq:betaxeq}-\eqref{eq:chemgradeq}} predict the  total Hall voltage \cite{supp}
\begin{align}
\label{eq:WeakBVy} 
\Delta  V_{\rm tot} & =  \frac{   \mathrm{e} l_{\rm imp} E_\mathrm{x}}{ m_{\rm eff} v_{\rm F} } \, \Big[  l_{\rm G} \,  \Big( \frac{m_{\rm eff} v_{\rm F} \eta_{\rm H}}{ \mathrm{e}  \eta l_{\rm imp} }-B \Big) \\
& \times \left( A_1 
\sinh \! \left(W l_{\rm G}^{-1}\right) + A_2 \left[\cosh \! \left(W l_{\rm G}^{-1} \right) \! - \! 1\right] \right)  \! -\!  BW \Big], \nonumber
\end{align}
where the functions $A_{1,2}(l_\mathrm{s},W)$ are explicitly defined in Ref.~\cite{supp}. For $W/l_{\rm G}\ll 1$, corresponding to dominant shear effects, Eq.~\eqref{eq:WeakBVy} implies up to second order in $l_\mathrm{G}^{-1}$ 
\begin{align} \label{eq:HallVoltagesCase3}
\Delta 	V_{\eta_{\rm H}} &=  \frac{\eta_\mathrm{H} }{\eta}  E_\mathrm{x} \left[ W -  
  { 1  \over  12    l_{\rm G}^2 }
 \left( W^3 \! + \! 6 l_{\rm s} W^2 \! + \! 6 l_{\rm s}^2 W\right) \right] ,\nonumber \\
\Delta 	V_{B} &= - \frac{\mathrm{sgn}(B) \, E_\mathrm{x}}{ 3 r_\mathrm{c} l_\mathrm{ee}} \left( W^3 \! + \! 6 l_{\rm s} W^2 \! + \! 6 l_{\rm s}^2 W\right) \ .
\end{align}
Hence, for weak magnetic fields, this implies the ratio
\begin{align} \label{ratioweak}
\dfrac{\vert \Delta V_{\eta_\mathrm{_H}} \vert}{ \vert \Delta V_{\rm B} \vert} = \dfrac{l_\mathrm{ee}^2}{\frac{1}{6} W^2 + l_\mathrm{s}W+l_\mathrm{s}^2} + 2 \dfrac{l_\mathrm{ee}}{l_\mathrm{imp}} \ .
\end{align}

\noindent
Thus, for small magnetic fields and $W/l_\mathrm{G} \! \ll \! 1$, this ratio is completely determined by the characteristic length scales of the system. Its modulus decreases as a function of $W$, $l_\mathrm{s}$ and $l_\mathrm{imp}$, whereas it increases with $l_\mathrm{ee}$. While the hydrodynamic assumption $l_\mathrm{ee} \! \ll \! l_\mathrm{imp}$ strongly suppresses the second term in Eq.~\eqref{ratioweak}  \footnote{
In this limit, according to Eq.~\eqref{eq:HallVoltagesCase3}, $\Delta  V_{\eta_\mathrm{H}}$ becomes insensitive to the concrete form of the boundary conditions.}, 
the first term experimentally can  be much larger than one \cite{Gusev18,Gusev18:2,Bakarov18}.
For very small $l_\mathrm{s}$ and $W$ in comparison to $l_\mathrm{ee}$, engineered e.g. in Ref.~\cite{Moessner19}, the Hall viscous contribution can strongly dominate the Lorentz signal.
In addition, Eq.~\eqref{eq:HallVoltagesCase3}  provides the density dependence of each voltage contribution.  For temperatures much smaller than the Fermi energy, we obtain the dependence  \cite{PhysRevB.26.4421,bookref}
\begin{align} \label{eq:densitydep}
	\Delta  V_{\eta_\mathrm{H}} & = f_1 \, \rho_0 + f_2(n_\mathrm{imp})  \quad  \wedge \quad
\Delta V_B = f_3 \, \rho_0^{-2}  \ , 
\end{align}
where $n_\mathrm{imp}$ is the impurity concentration and $f_{1,2,3}$ are density independent functions, given in Ref.~\cite{supp}. Explicitly, Eq.~\eqref{eq:densitydep} predicts 
$ \Delta V_{\eta_\mathrm{_H}} \! / \Delta  V_{\rm B} \! = \! \mathcal{O}_\mathrm{clean}(\rho_0^3) + \mathcal{O}_\mathrm{imp}(\rho_0^2)$. Hence, in the weak field limit and for $W/l_\mathrm{G} \! \ll \! 1$, the Hall viscous contribution becomes strongly enhanced in comparison to the Lorentz force signal as the carrier density increases. 
Summarizing, the distinct dependence of $\Delta V_{\eta_\mathrm{H}}$ and $ \Delta V_B$ on $\rho_0, W, l_\mathrm{s},l_\mathrm{imp}$ and $l_\mathrm{ee}$ allows to experimentally distinguish these two contributions in the limit of weak magnetic fields and clean systems. Beyond this limit, the Hall viscous and Lorentz force contributions to $\Delta V_\mathrm{tot}$
need to be evaluated numerically. By solving Eqs.~\eqref{eq:betaxeq} and \eqref{eq:chemgradeq} for the velocity profile, we obtain
\begin{align} \label{eq:intermediate}
\! \! \Delta   V_B \!  = -\! B \! \int_0^W \! \! \! \!\mathrm{d} y \,  v_\mathrm{x} (y) \  \!  & \wedge \!  \  \Delta   V_{\eta_\mathrm{H}} \!  = \!  \frac{\eta_\mathrm{H} }{ {\rm e} \rho_0} \!\left[\partial_\mathrm{y} v_\mathrm{x}(y)\right]_{y=0}^W .
\end{align}
In what follows, we  numerically investigate the dependence of these voltage contributions on $B,W,l_\mathrm{s},l_\mathrm{imp}$ and $l_\mathrm{ee}$.
Notice, that while $\Delta V_B$ is proportional to the integrated value of $v_\mathrm{x}(y)$, which characterizes the overall fluid momentum, $\Delta V_{\eta_\mathrm{H}}$ is totally determined by the gradient of the velocity profile at the boundaries of the channel. 
In Fig.~\ref{fig:poiseuille}, we plot the velocity profile of a general Fermi liquid for different magnetic fields. 
With increasing $\vert B \vert$ the fluid is accelerated due to the corresponding  suppression of $\eta$ (cf.~Eq.~\eqref{eq:eta}). According to Eq.~\eqref{eq:intermediate}, this leads to an enhancement of $\vert \Delta V_B \vert$. For large fields, $\eta$ vanishes and solely impurity scattering is responsible for momentum relaxation, leading to an Ohmic velocity profile  \cite{supp}.

\vspace{.2cm}
%
%
%
%
%
%
%
%
The dependence of $\Delta V_{\eta_\mathrm{H}}$ on the magnetic field is more subtle. Figure~\ref{fig:poiseuille} shows that for weak magnetic fields, the gradient $\vert \partial_\mathrm{y} v_\mathrm{x}(y) \vert_{y=0,W} \vert$ increases as a function of $ \vert B \vert $. According to Eq.~\eqref{eq:intermediate}, this corresponds to an enhancement of $\vert \Delta V_{\eta_\mathrm{H}} \vert$. As discussed above, impurity scattering causes an Ohmic (flat) velocity profile for large magnetic fields. This  decreases  $\vert \partial_\mathrm{y} v_\mathrm{x}(y) \vert_{y=0,W} \vert$ and therefore reduces $ \vert \Delta V_{\eta_\mathrm{H}} \vert$. 
Altogether, this implies that systems in which Hall viscous effects dominate the weak magnetic field regime  are eventually always driven to  $\vert \Delta V_{\eta_{\rm H}} \vert / \vert \Delta V_{\rm B} \vert \! \ll \! 1$. The transition from Hall viscous to Lorentz force dominated transport occurs for $\Delta V_{\eta_{\rm H}}/ \Delta V_{\rm B} \! = \! -1$, where $\Delta V_\mathrm{tot}=0$. This happens at a certain critical magnetic field $B_\mathrm{c}$, which strongly depends on $W$, $l_\mathrm{s}$ and $l_\mathrm{ee}$.
\begin{figure}[t!]
\flushleft{
\includegraphics[scale=0.125]{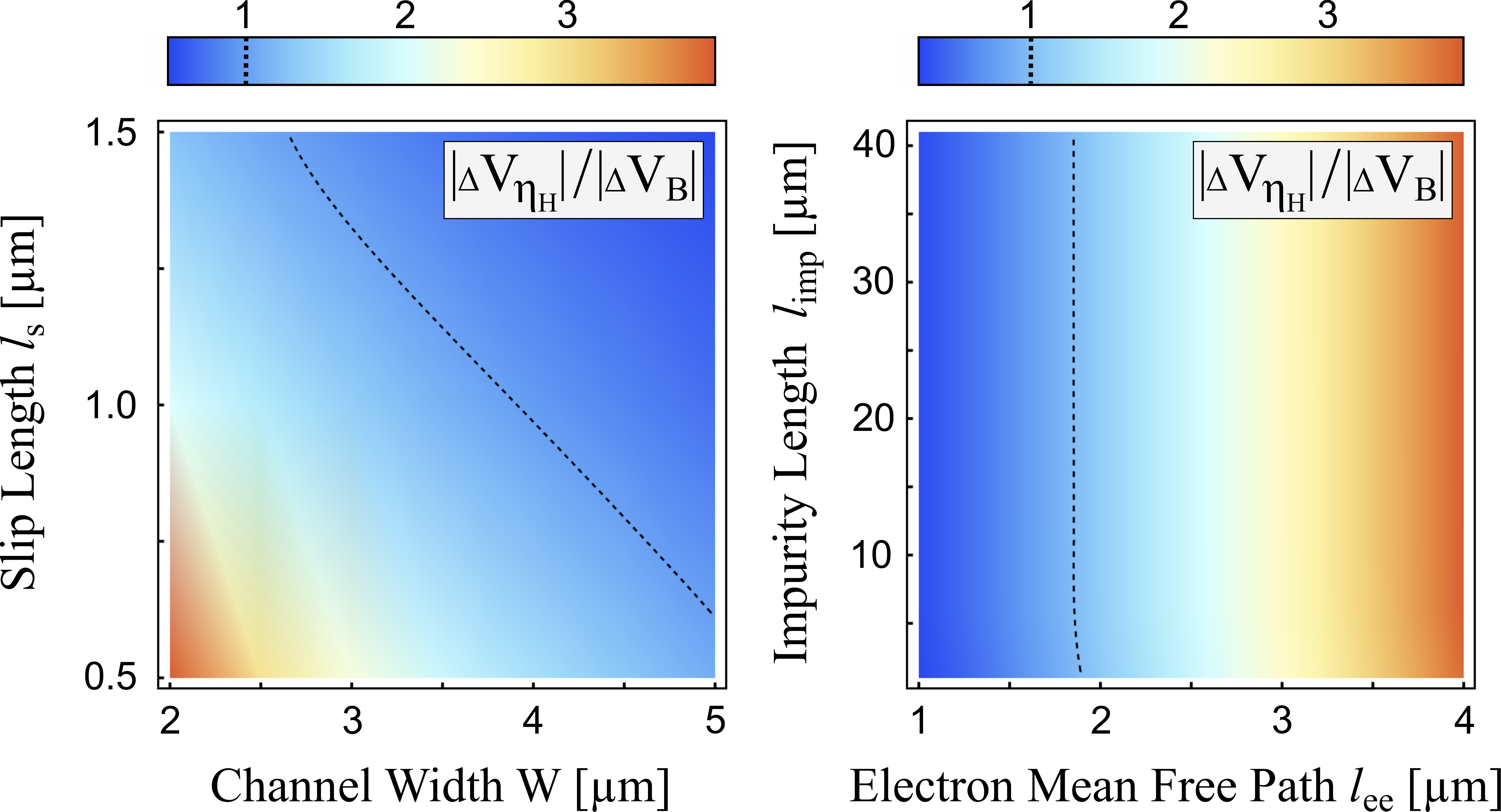}}
\caption{$ \vert \Delta V_{\eta_{\rm H}} \vert / \vert  \Delta V_{\rm B} \vert$ for different fluids at $B\!~=~\! 10$mT. In the left (right) panel, this ratio is shown  as a function of $W$ vs. $l_\mathrm{s}$ ($l_\mathrm{ee}$ vs. $l_\mathrm{imp}$). For those parameters that are not altered, we choose $l_\mathrm{s} \!=\! 0.5 {\rm \mu m}$, $ W\!=\!3\mu \rm m$, $\rho_0\!=\!9.1\times 10^{11} \rm cm^{-2}$, $T\!=\!1.4 \rm K$, $\eta_0 \! = \! 1.7 \!\times \! 10^{-16} \rm{J s/m^2}$ and $l_{\rm imp} \! = \! 40 \mu \rm m$ \cite{Gusev18}. While $\vert \Delta V_{\eta_{\rm H}} \vert / \vert \Delta V_{\rm B} \vert$ strongly decreases with $W$ and $l_\mathrm{s}$, the ratio is rather unaffected by $l_\mathrm{imp}$ and increases  as a function of $l_\mathrm{ee}$. The black dashed curve, for which  $\Delta V_{\eta_{\rm H}}/ \Delta V_{\rm B} \! = \! -1 $, shows the range of parameters where the total Hall voltage vanishes. \label{fig:vhall}}
\end{figure}
This is shown in Fig.~\ref{fig:vhall}, which
displays $\vert \Delta V_{\eta_{\rm H}} \vert / \vert \Delta V_{\rm B} \vert $ as a function of 
$W,l_\mathrm{s},l_\mathrm{ee}$ and $l_\mathrm{imp}$ \footnote[10]{Note that the system lies within the range of validity of the hydrodynamic regime, even though  $l_{\rm ee}\lesssim W $ \cite{2019arXiv190511662S}.}.
According to Eq.~\eqref{eq:intermediate},
the dependence of this ratio on these parameters can be explained by analyzing the corresponding velocity profiles, depicted in Fig. \ref{fig:poiseuille}. As $l_{\rm s}$ increases, the gradient $\vert \partial_\mathrm{y} v_\mathrm{x}(y) \vert_{y=0,W} \vert$ decreases, leading to a reduction of $\vert \Delta V_{\eta_{\rm H}} \vert$. In contrast, since the overall fluid momentum  is enhanced,  $ \vert \Delta V_{B} \vert$ increases as a function of  the slip length. Therefore,  $ \vert \Delta V_{\eta_{\rm H}} \vert/ \vert \Delta V_{\rm B} \vert $ decreases with  $l_\mathrm{s}$, as illustrated in Fig.~\ref{fig:vhall}. Increasing the channel width leads to the same result, since it also increases the overall fluid momentum and at the same time decreases $\vert \partial_\mathrm{y} v_\mathrm{x}(y) \vert_{y=0,W} \vert$.  In contrast, $\vert \Delta V_{\eta_{\rm H}} \vert / \vert \Delta V_{\rm B} \vert $ increases  with the electron-electron mean free path. According to Eq.~\eqref{eq:eta}, the viscosities  $\eta$, $\vert \eta_\mathrm{H} \vert$ as well as the ratio $\vert \eta_\mathrm{H} \vert /\eta$ increase as a function of $l_\mathrm{ee}$ as long as $l_\mathrm{ee} \! \ll \! r_\mathrm{c}$, implied by the hydrodynamic assupmtion. This reduces the overall fluid momentum but increases $\vert \Delta V_{\eta_{\rm H}} \vert $. 
Last but not least, Fig.~\ref{fig:vhall} shows that since $l_\mathrm{ee} \ll l_\mathrm{imp}$, the impurity length does not significantly change  $\Delta V_{\eta_{\rm H}}/ \Delta V_{\rm B}$. 

\begin{figure}[t!]
\includegraphics[scale=0.125]{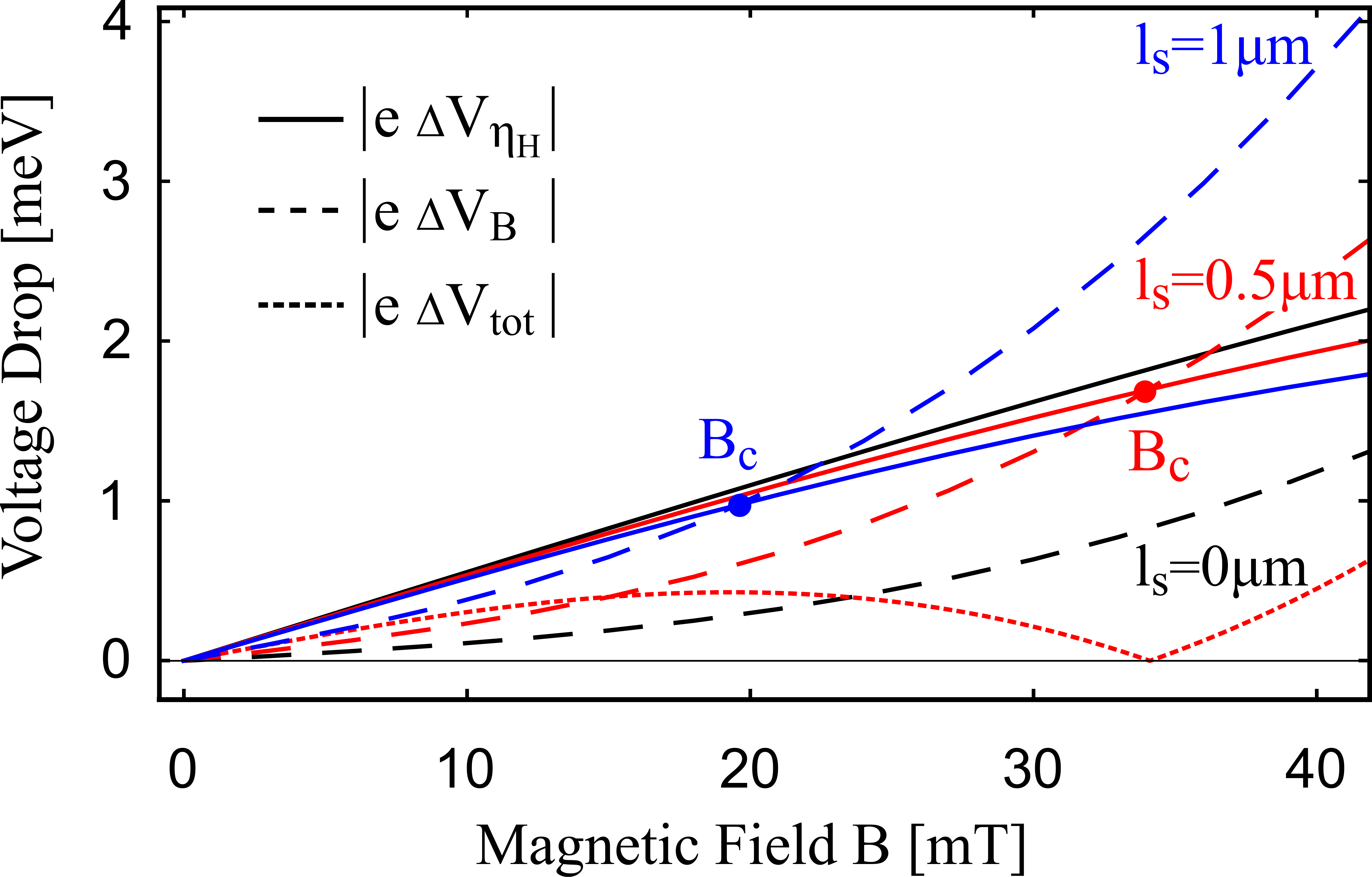}
\caption{Absolute values of the Lorentz $\Delta V_B$ and Hall viscous contribution $\Delta V_{\eta_\mathrm{H}}$ to the total Hall voltage $\Delta V_\mathrm{tot}$ in GaAs are shown as functions of the magnetic field $B$ for \mbox{$l_{\rm s} = 0, 0.5, 1.0 {\rm \mu m}$.} Parameters for this calculation are given in the caption of Fig.~\ref{fig:vhall}. \mbox{For $B < B_\mathrm{c}$}, we find $\vert \Delta V_{\eta_{\rm H}} \vert / \vert \Delta V_{B}\vert > 1$, whereas otherwise $ \vert \Delta V_{\eta_{\rm H}} \vert / \vert \Delta V_{B} \vert < 1  $. At $B=B_\mathrm{c}$, the ratio $\Delta V_{\eta_{\rm H}}/ \Delta V_{\rm B} \! = \! -1 $ implying a vanishing Hall voltage $\Delta V_\mathrm{tot} \! = \! 0$.  
 \label{fig:gaas1}} 
\end{figure}

\noindent
To demonstrate the experimental relevance of our predictions, Fig.~\ref{fig:gaas1} finally shows  $\vert \Delta V_{\eta_{\rm H}} \vert$, $ \vert \Delta V_{\rm B} \vert$
and $\vert \Delta V_{\rm tot} \vert$ in  GaAs  as a function of the magnetic field  \cite{Gusev18,Gusev18:2}\footnotemark[10].
Since in this material the slip length is highly dependent on the etching technique, we plot the corresponding curves for $l_{\rm s} = 0, 0.5, 1.0 {\rm \mu  m}$ \footnote{Private communication with Prof. Laurens Molenkamp, University of Wuerzburg}.
For $\vert B \vert \! \ll \! \vert B_\mathrm{c} \vert$, the  ratio $\vert \Delta V_{\eta_{\rm H}} \vert / \vert \Delta V_{\rm B} \vert \gg 1$, giving rise to a Hall viscosity dominated transport regime. As indicated by Eq.~\eqref{ratioweak} and confirmed by Fig.~\ref{fig:vhall}, this property is strongly pronounced if the slip length decreases.
With the exception of $l_\mathrm{ee}$, it is possible to evaluate all parameters in 
Eqs.~\!\!\eqref{eq:betaxeq}-\eqref{eq:eta} precisely. Therefore, measuring $\Delta V_{\eta_{\rm H}}$ and $\Delta V_{\rm B}$ with increasing $\vert B \vert$, enables us to determine the electron-electron mean free path in this sample by fitting theoretical and experimental data. In general, this procedure works for any Fermi liquid and therefore is  a powerful tool to evaluate the electron interaction strength. For $B \! = \! B_\mathrm{c}$, the two voltage contributions are equal in magnitude but opposite in sign, leading to $\Delta V_\mathrm{tot}=0$. Experimentally, this smoking gun feature can be used to prove our Hall viscous magnetotransport theory.  For $ \vert B \vert \! \gg \! \vert B_\mathrm{c} \vert$,  Fig.~\ref{fig:gaas1} shows that transport is dominated by the Lorentz signal. Eventually, this leads to an Ohmic, impurity driven velocity profile \cite{supp}. 
Our present approach does not incorporate the formation of Landau levels which in GaAs occurs beyond the applicability of hydrodynamics for $\vert B \vert \! \gtrsim \! 1$T.  
Finally, we remark that the Hall voltage we predict depends non-linearly on the magnetic field strength. Since ballistic transport predicts $\Delta V_{\rm B} \propto B$, this non-linear behavior is a direct consequence of the hydrodynamic regime.

\vspace{.2cm}

\textit{Local Hall response.}
So far, we focused on global voltage measurements across the entire channel. In addition,
Eq.~\eqref{eq:intermediate} predicts the local voltage contributions:  
\begin{align}
\label{eq:voltages referee A}
\Delta 	V_{\eta_{\rm H}}  \! &=  \! \frac{\eta_\mathrm{H} }{\eta}  E_\mathrm{x} \left[ y \! - \!  
  { 1  \over  12     l_{\rm G}^2 }
 \left[6l_s(l_s \! + \! W) \! + \!  y (3 W \! - \! 2 y)\right]y\right] \nonumber \\
\Delta 	V_{B} \! &= \! -  \frac{ \mathrm{sgn}(B)   E_\mathrm{x}}{3 r_\mathrm{c} l_\mathrm{ee}}\left[6l_s(l_s \! + \! W) +  y (3 W \! - \! 2 y)\right]y .
\end{align}
Hence, the total Hall voltage  $\Delta V_\mathrm{tot}$ scales as $y^3$, whereas the associated Hall field $E_{\rm y} \! = \! - \mathrm{d}\Delta V_{\rm tot}/ \mathrm{d} y$ has parabolic structure. For large $l_\mathrm{imp}$, the curvature of this parabola, $\kappa \! = \! -  4 E_\mathrm{x} \left[ \mathrm{sgn}(B)/r_\mathrm{c} + \eta_\mathrm{H} / ( \eta l_\mathrm{imp} ) \right] \! /l_\mathrm{ee} $, is mainly \mbox{characterized} by the Lorentz contribution. In contrast, the Hall viscous signal significantly defines the offset of this parabola, 
\begin{align}
\label{eq:HallFieldBoundary} E_\mathrm{y}(0) = E_\mathrm{x} \left[{2l_{\rm s}(l_{\rm s} \! + \! W) \over \mathrm{sgn}(B) r_\mathrm{c} l_\mathrm{ ee}} \! - \! \left( \! 1 \! - \! \dfrac{2 l_\mathrm{s}(  l_\mathrm{s} \!+ \! W)}{ l_\mathrm{ee} l_ \mathrm{imp}} \right) \! {\eta_{\rm H} \over \eta}\right]  .
\end{align}
For $l_s \! \ll \! l_\mathrm{ee} \! \ll \! l_\mathrm{imp}$, this relation becomes $E_\mathrm{y}(0) \! \simeq \! E_\mathrm{x} \eta_\mathrm{H} /\eta $.
Recently, the authors of Ref.~\cite{2019arXiv190511662S}
measured the local Hall voltage in the hydrodynamic regime of graphene, particularly satisfying $l_s \! \simeq \! l_\mathrm{ee} \! \ll \! W$. They explicitly reproduced 
the parabolic form of the Hall field, as well as that $\vert \kappa \vert$ decreases with increasing $l_\mathrm{ee}$. 

\vspace{.2cm}



\emph{Conclusions.} Our results provide new insights into  Hall viscous effects of two-dimensional non-relativistic  electron fluids in external magnetic fields. We present a setup that allows to distinguish unambiguously  between Hall viscous and Lorentz force contributions to the total Hall voltage. In particular, we prove that the  ratio of Hall viscous to Lorentz force induced voltage contributions is negative and that in clean systems its modulus decreases with the width, slip-length and carrier density, while it increases with the electron-electron mean free path of our setup.
Additionally, we show that this identification is possible by measuring the Hall voltage locally. In particular, we prove that the total Hall field is parabolic in the transverse channel coordinate, as it was recently measured in Ref.~\cite{2019arXiv190511662S}. While the curvature of this parabola mainly originates from Lorentz effects, its offset is characterized by the Hall viscosity. For typical Fermi liquids such as GaAs, we find that the Hall viscous signal dominates over the Lorentz force contribution up to a critical magnetic field  which is on the order of a few tens of millitesla.~At this field, the total Hall voltage vanishes, clearly indicating Hall viscous transport. Hence, our predictions pave the way to experimentally identify Hall viscous effects and explicitly demonstrate how to determine the electron-electron interaction strength. 
Possible future directions of our approach include two-dimensional massive Dirac systems, where parity and time-reversal symmetry are already broken for vanishing magnetic fields \cite{Tutschku19} and a torsional Hall viscous term  is present \cite{Hoyos:2014pba,Hughes12}.\\

\begin{acknowledgments}
\emph{Acknowledgments.} We thank Carlos Hoyos, Andrew Lucas, Hartmut Buhmann, Laurens Molenkamp and Igor Gornyi for useful discussions. We acknowledge financial support through the Deutsche Forschungsgemeinschaft (DFG, German Research Foundation), project-id 258499086 - SFB 1170 'ToCoTronics', the ENB Graduate School on Topological Insulators and through the W\"urzburg-Dresden Cluster of Excellence on Complexity and Topology in Quantum Matter - ct.qmat \mbox{(EXC 2147, project-id 39085490)}.
\end{acknowledgments}

\vspace{-.2cm}

\bibliographystyle{apsrev4-1}
\bibliography{biblio2}

\begin{thebibliography}{43}%
\makeatletter
\providecommand \@ifxundefined [1]{%
 \@ifx{#1\undefined}
}%
\providecommand \@ifnum [1]{%
 \ifnum #1\expandafter \@firstoftwo
 \else \expandafter \@secondoftwo
 \fi
}%
\providecommand \@ifx [1]{%
 \ifx #1\expandafter \@firstoftwo
 \else \expandafter \@secondoftwo
 \fi
}%
\providecommand \natexlab [1]{#1}%
\providecommand \enquote  [1]{``#1''}%
\providecommand \bibnamefont  [1]{#1}%
\providecommand \bibfnamefont [1]{#1}%
\providecommand \citenamefont [1]{#1}%
\providecommand \href@noop [0]{\@secondoftwo}%
\providecommand \href [0]{\begingroup \@sanitize@url \@href}%
\providecommand \@href[1]{\@@startlink{#1}\@@href}%
\providecommand \@@href[1]{\endgroup#1\@@endlink}%
\providecommand \@sanitize@url [0]{\catcode `\\12\catcode `\$12\catcode
  `\&12\catcode `\#12\catcode `\^12\catcode `\_12\catcode `\%12\relax}%
\providecommand \@@startlink[1]{}%
\providecommand \@@endlink[0]{}%
\providecommand \url  [0]{\begingroup\@sanitize@url \@url }%
\providecommand \@url [1]{\endgroup\@href {#1}{\urlprefix }}%
\providecommand \urlprefix  [0]{URL }%
\providecommand \Eprint [0]{\href }%
\providecommand \doibase [0]{http://dx.doi.org/}%
\providecommand \selectlanguage [0]{\@gobble}%
\providecommand \bibinfo  [0]{\@secondoftwo}%
\providecommand \bibfield  [0]{\@secondoftwo}%
\providecommand \translation [1]{[#1]}%
\providecommand \BibitemOpen [0]{}%
\providecommand \bibitemStop [0]{}%
\providecommand \bibitemNoStop [0]{.\EOS\space}%
\providecommand \EOS [0]{\spacefactor3000\relax}%
\providecommand \BibitemShut  [1]{\csname bibitem#1\endcsname}%
\let\auto@bib@innerbib\@empty
\bibitem [{\citenamefont {Molenkamp}\ and\ \citenamefont
  {de~Jong}(1994{\natexlab{a}})}]{Molenkamp:1994ii}%
  \BibitemOpen
  \bibfield  {author} {\bibinfo {author} {\bibfnamefont {L.~W.}\ \bibnamefont
  {Molenkamp}}\ and\ \bibinfo {author} {\bibfnamefont {M.~J.~M.}\ \bibnamefont
  {de~Jong}},\ }\href@noop {} {\bibfield  {journal} {\bibinfo  {journal} {Phys.
  Rev. B}\ }\textbf {\bibinfo {volume} {49}},\ \bibinfo {pages} {5038}
  (\bibinfo {year} {1994}{\natexlab{a}})}\BibitemShut {NoStop}%
\bibitem [{\citenamefont {Molenkamp}\ and\ \citenamefont
  {de~Jong}(1994{\natexlab{b}})}]{Molenkamp:1994kb}%
  \BibitemOpen
  \bibfield  {author} {\bibinfo {author} {\bibfnamefont {L.~W.}\ \bibnamefont
  {Molenkamp}}\ and\ \bibinfo {author} {\bibfnamefont {M.~J.~M.}\ \bibnamefont
  {de~Jong}},\ }\href@noop {} {\bibfield  {journal} {\bibinfo  {journal}
  {Solid-State Electronics}\ }\textbf {\bibinfo {volume} {37}},\ \bibinfo
  {pages} {551} (\bibinfo {year} {1994}{\natexlab{b}})}\BibitemShut {NoStop}%
\bibitem [{\citenamefont {{de Jong}}\ and\ \citenamefont
  {{Molenkamp}}(1995)}]{deJong:1995bn}%
  \BibitemOpen
  \bibfield  {author} {\bibinfo {author} {\bibfnamefont {M.~J.~M.}\
  \bibnamefont {{de Jong}}}\ and\ \bibinfo {author} {\bibfnamefont {L.~W.}\
  \bibnamefont {{Molenkamp}}},\ }\href {\doibase 10.1103/PhysRevB.51.13389}
  {\bibfield  {journal} {\bibinfo  {journal} {Phys. Rev. B}\ }\textbf {\bibinfo
  {volume} {51}},\ \bibinfo {pages} {13389} (\bibinfo {year}
  {1995})}\BibitemShut {NoStop}%
\bibitem [{\citenamefont {Lucas}\ and\ \citenamefont {Fong}(2018)}]{Lucas2018}%
  \BibitemOpen
  \bibfield  {author} {\bibinfo {author} {\bibfnamefont {A.}~\bibnamefont
  {Lucas}}\ and\ \bibinfo {author} {\bibfnamefont {K.~C.}\ \bibnamefont
  {Fong}},\ }\href {\doibase 10.1088/1361-648x/aaa274} {\bibfield  {journal}
  {\bibinfo  {journal} {Journal of Physics: Condensed Matter}\ }\textbf
  {\bibinfo {volume} {30}},\ \bibinfo {pages} {053001} (\bibinfo {year}
  {2018})}\BibitemShut {NoStop}%
\bibitem [{\citenamefont {Bandurin}\ \emph {et~al.}(2016)\citenamefont
  {Bandurin}, \citenamefont {Torre}, \citenamefont {Kumar}, \citenamefont
  {Shalom}, \citenamefont {Tomadin}, \citenamefont {Principi}, \citenamefont
  {Auton}, \citenamefont {Khestanova}, \citenamefont {Novoselov}, \citenamefont
  {Grigorieva}, \citenamefont {Ponomarenko}, \citenamefont {Geim},\ and\
  \citenamefont {Polini}}]{bandurin2016negative}%
  \BibitemOpen
  \bibfield  {author} {\bibinfo {author} {\bibfnamefont {D.}~\bibnamefont
  {Bandurin}}, \bibinfo {author} {\bibfnamefont {I.}~\bibnamefont {Torre}},
  \bibinfo {author} {\bibfnamefont {R.~K.}\ \bibnamefont {Kumar}}, \bibinfo
  {author} {\bibfnamefont {M.~B.}\ \bibnamefont {Shalom}}, \bibinfo {author}
  {\bibfnamefont {A.}~\bibnamefont {Tomadin}}, \bibinfo {author} {\bibfnamefont
  {A.}~\bibnamefont {Principi}}, \bibinfo {author} {\bibfnamefont
  {G.}~\bibnamefont {Auton}}, \bibinfo {author} {\bibfnamefont
  {E.}~\bibnamefont {Khestanova}}, \bibinfo {author} {\bibfnamefont
  {K.}~\bibnamefont {Novoselov}}, \bibinfo {author} {\bibfnamefont
  {I.}~\bibnamefont {Grigorieva}}, \bibinfo {author} {\bibfnamefont
  {L.}~\bibnamefont {Ponomarenko}}, \bibinfo {author} {\bibfnamefont
  {A.}~\bibnamefont {Geim}}, \ and\ \bibinfo {author} {\bibfnamefont
  {M.}~\bibnamefont {Polini}},\ }\href@noop {} {\bibfield  {journal} {\bibinfo
  {journal} {Science}\ }\textbf {\bibinfo {volume} {351}},\ \bibinfo {pages}
  {1055} (\bibinfo {year} {2016})}\BibitemShut {NoStop}%
\bibitem [{\citenamefont {Krishna~Kumar}\ \emph {et~al.}(2017)\citenamefont
  {Krishna~Kumar}, \citenamefont {Bandurin}, \citenamefont {Pellegrino},
  \citenamefont {Cao}, \citenamefont {Principi}, \citenamefont {Guo},
  \citenamefont {Auton}, \citenamefont {Ben~Shalom}, \citenamefont
  {Ponomarenko}, \citenamefont {Falkovich}, \citenamefont {Watanabe},
  \citenamefont {Taniguchi}, \citenamefont {Grigorieva}, \citenamefont
  {Levitov}, \citenamefont {Polini},\ and\ \citenamefont
  {Geim}}]{KrishnaKumar2017}%
  \BibitemOpen
  \bibfield  {author} {\bibinfo {author} {\bibfnamefont {R.}~\bibnamefont
  {Krishna~Kumar}}, \bibinfo {author} {\bibfnamefont {D.~A.}\ \bibnamefont
  {Bandurin}}, \bibinfo {author} {\bibfnamefont {F.~M.~D.}\ \bibnamefont
  {Pellegrino}}, \bibinfo {author} {\bibfnamefont {Y.}~\bibnamefont {Cao}},
  \bibinfo {author} {\bibfnamefont {A.}~\bibnamefont {Principi}}, \bibinfo
  {author} {\bibfnamefont {H.}~\bibnamefont {Guo}}, \bibinfo {author}
  {\bibfnamefont {G.~H.}\ \bibnamefont {Auton}}, \bibinfo {author}
  {\bibfnamefont {M.}~\bibnamefont {Ben~Shalom}}, \bibinfo {author}
  {\bibfnamefont {L.~A.}\ \bibnamefont {Ponomarenko}}, \bibinfo {author}
  {\bibfnamefont {G.}~\bibnamefont {Falkovich}}, \bibinfo {author}
  {\bibfnamefont {K.}~\bibnamefont {Watanabe}}, \bibinfo {author}
  {\bibfnamefont {T.}~\bibnamefont {Taniguchi}}, \bibinfo {author}
  {\bibfnamefont {I.~V.}\ \bibnamefont {Grigorieva}}, \bibinfo {author}
  {\bibfnamefont {L.~S.}\ \bibnamefont {Levitov}}, \bibinfo {author}
  {\bibfnamefont {M.}~\bibnamefont {Polini}}, \ and\ \bibinfo {author}
  {\bibfnamefont {A.~K.}\ \bibnamefont {Geim}},\ }\href
  {https://doi.org/10.1038/nphys4240} {\bibfield  {journal} {\bibinfo
  {journal} {Nature Physics}\ }\textbf {\bibinfo {volume} {13}},\ \bibinfo
  {pages} {1182} (\bibinfo {year} {2017})}\BibitemShut {NoStop}%
\bibitem [{\citenamefont {{Moll}}\ \emph {et~al.}(2016)\citenamefont {{Moll}},
  \citenamefont {{Kushwaha}}, \citenamefont {{Nandi}}, \citenamefont
  {{Schmidt}},\ and\ \citenamefont {{Mackenzie}}}]{Moll:2016ju}%
  \BibitemOpen
  \bibfield  {author} {\bibinfo {author} {\bibfnamefont {P.~J.~W.}\
  \bibnamefont {{Moll}}}, \bibinfo {author} {\bibfnamefont {P.}~\bibnamefont
  {{Kushwaha}}}, \bibinfo {author} {\bibfnamefont {N.}~\bibnamefont {{Nandi}}},
  \bibinfo {author} {\bibfnamefont {B.}~\bibnamefont {{Schmidt}}}, \ and\
  \bibinfo {author} {\bibfnamefont {A.~P.}\ \bibnamefont {{Mackenzie}}},\
  }\href {\doibase 10.1126/science.aac8385} {\bibfield  {journal} {\bibinfo
  {journal} {Science}\ }\textbf {\bibinfo {volume} {351}},\ \bibinfo {pages}
  {1061} (\bibinfo {year} {2016})}\BibitemShut {NoStop}%
\bibitem [{\citenamefont {Gusev}\ \emph {et~al.}(2018)\citenamefont {Gusev},
  \citenamefont {Levin}, \citenamefont {Levinson},\ and\ \citenamefont
  {Bakarov}}]{Gusev18}%
  \BibitemOpen
  \bibfield  {author} {\bibinfo {author} {\bibfnamefont {G.~M.}\ \bibnamefont
  {Gusev}}, \bibinfo {author} {\bibfnamefont {A.~D.}\ \bibnamefont {Levin}},
  \bibinfo {author} {\bibfnamefont {E.~V.}\ \bibnamefont {Levinson}}, \ and\
  \bibinfo {author} {\bibfnamefont {A.~K.}\ \bibnamefont {Bakarov}},\ }\href
  {\doibase 10.1103/PhysRevB.98.161303} {\bibfield  {journal} {\bibinfo
  {journal} {Phys. Rev. B}\ }\textbf {\bibinfo {volume} {98}},\ \bibinfo
  {pages} {161303} (\bibinfo {year} {2018})}\BibitemShut {NoStop}%
\bibitem [{\citenamefont {{Gusev}}\ \emph {et~al.}(2018)\citenamefont
  {{Gusev}}, \citenamefont {{Levin}}, \citenamefont {{Levinson}},\ and\
  \citenamefont {{Bakarov}}}]{Gusev18:2}%
  \BibitemOpen
  \bibfield  {author} {\bibinfo {author} {\bibfnamefont {G.~M.}\ \bibnamefont
  {{Gusev}}}, \bibinfo {author} {\bibfnamefont {A.~D.}\ \bibnamefont
  {{Levin}}}, \bibinfo {author} {\bibfnamefont {E.~V.}\ \bibnamefont
  {{Levinson}}}, \ and\ \bibinfo {author} {\bibfnamefont {A.~K.}\ \bibnamefont
  {{Bakarov}}},\ }\href@noop {} {\bibfield  {journal} {\bibinfo  {journal} {AIP
  Advances}\ }\textbf {\bibinfo {volume} {8}},\ \bibinfo {pages} {025318}
  (\bibinfo {year} {2018})}\BibitemShut {NoStop}%
\bibitem [{\citenamefont {Levin}\ \emph {et~al.}(2018)\citenamefont {Levin},
  \citenamefont {Gusev}, \citenamefont {Levinson}, \citenamefont {Kvon},\ and\
  \citenamefont {Bakarov}}]{Bakarov18}%
  \BibitemOpen
  \bibfield  {author} {\bibinfo {author} {\bibfnamefont {A.~D.}\ \bibnamefont
  {Levin}}, \bibinfo {author} {\bibfnamefont {G.~M.}\ \bibnamefont {Gusev}},
  \bibinfo {author} {\bibfnamefont {E.~V.}\ \bibnamefont {Levinson}}, \bibinfo
  {author} {\bibfnamefont {Z.~D.}\ \bibnamefont {Kvon}}, \ and\ \bibinfo
  {author} {\bibfnamefont {A.~K.}\ \bibnamefont {Bakarov}},\ }\href {\doibase
  10.1103/PhysRevB.97.245308} {\bibfield  {journal} {\bibinfo  {journal} {Phys.
  Rev. B}\ }\textbf {\bibinfo {volume} {97}},\ \bibinfo {pages} {245308}
  (\bibinfo {year} {2018})}\BibitemShut {NoStop}%
\bibitem [{\citenamefont {{Avron}}\ \emph {et~al.}(1995)\citenamefont
  {{Avron}}, \citenamefont {{Seiler}},\ and\ \citenamefont
  {{Zograf}}}]{AvronSeilerZograf}%
  \BibitemOpen
  \bibfield  {author} {\bibinfo {author} {\bibfnamefont {J.~E.}\ \bibnamefont
  {{Avron}}}, \bibinfo {author} {\bibfnamefont {R.}~\bibnamefont {{Seiler}}}, \
  and\ \bibinfo {author} {\bibfnamefont {P.~G.}\ \bibnamefont {{Zograf}}},\
  }\href {\doibase 10.1103/PhysRevLett.75.697} {\bibfield  {journal} {\bibinfo
  {journal} {Phys. Rev. Lett.}\ }\textbf {\bibinfo {volume} {75}},\ \bibinfo
  {pages} {697} (\bibinfo {year} {1995})}\BibitemShut {NoStop}%
\bibitem [{\citenamefont {Hoyos}(2014)}]{Hoyos:2014pba}%
  \BibitemOpen
  \bibfield  {author} {\bibinfo {author} {\bibfnamefont {C.}~\bibnamefont
  {Hoyos}},\ }\href@noop {} {\bibfield  {journal} {\bibinfo  {journal} {Int. J.
  Mod. Phys. B}\ }\textbf {\bibinfo {volume} {28}},\ \bibinfo {pages} {1430007}
  (\bibinfo {year} {2014})}\BibitemShut {NoStop}%
\bibitem [{\citenamefont {Alekseev}(2016)}]{alekseev2016negative}%
  \BibitemOpen
  \bibfield  {author} {\bibinfo {author} {\bibfnamefont {P.}~\bibnamefont
  {Alekseev}},\ }\href@noop {} {\bibfield  {journal} {\bibinfo  {journal}
  {Phys. Rev. Lett.}\ }\textbf {\bibinfo {volume} {117}},\ \bibinfo {pages}
  {166601} (\bibinfo {year} {2016})}\BibitemShut {NoStop}%
\bibitem [{\citenamefont {Link}\ \emph {et~al.}(2018)\citenamefont {Link},
  \citenamefont {Narozhny}, \citenamefont {Kiselev},\ and\ \citenamefont
  {Schmalian}}]{Schmalian17}%
  \BibitemOpen
  \bibfield  {author} {\bibinfo {author} {\bibfnamefont {J.~M.}\ \bibnamefont
  {Link}}, \bibinfo {author} {\bibfnamefont {B.~N.}\ \bibnamefont {Narozhny}},
  \bibinfo {author} {\bibfnamefont {E.~I.}\ \bibnamefont {Kiselev}}, \ and\
  \bibinfo {author} {\bibfnamefont {J.}~\bibnamefont {Schmalian}},\ }\href
  {\doibase 10.1103/PhysRevLett.120.196801} {\bibfield  {journal} {\bibinfo
  {journal} {Phys. Rev. Lett.}\ }\textbf {\bibinfo {volume} {120}},\ \bibinfo
  {pages} {196801} (\bibinfo {year} {2018})}\BibitemShut {NoStop}%
\bibitem [{\citenamefont {Pena-Benitez}\ \emph {et~al.}(2019)\citenamefont
  {Pena-Benitez}, \citenamefont {Saha},\ and\ \citenamefont
  {Surowka}}]{Surowka19}%
  \BibitemOpen
  \bibfield  {author} {\bibinfo {author} {\bibfnamefont {F.}~\bibnamefont
  {Pena-Benitez}}, \bibinfo {author} {\bibfnamefont {K.}~\bibnamefont {Saha}},
  \ and\ \bibinfo {author} {\bibfnamefont {P.}~\bibnamefont {Surowka}},\
  }\href@noop {} {\bibfield  {journal} {\bibinfo  {journal} {Phys. Rev. B}\
  }\textbf {\bibinfo {volume} {99}},\ \bibinfo {pages} {045141} (\bibinfo
  {year} {2019})}\BibitemShut {NoStop}%
\bibitem [{\citenamefont {Copetti}\ and\ \citenamefont
  {Landsteiner}(2019)}]{Landsteiner19}%
  \BibitemOpen
  \bibfield  {author} {\bibinfo {author} {\bibfnamefont {C.}~\bibnamefont
  {Copetti}}\ and\ \bibinfo {author} {\bibfnamefont {K.}~\bibnamefont
  {Landsteiner}},\ }\href@noop {} {\bibfield  {journal} {\bibinfo  {journal}
  {arXiv:1901.11403}\ } (\bibinfo {year} {2019})}\BibitemShut {NoStop}%
\bibitem [{\citenamefont {Berdyugin}\ \emph {et~al.}(2019)\citenamefont
  {Berdyugin}, \citenamefont {Xu}, \citenamefont {Pellegrino}, \citenamefont
  {Krishna~Kumar}, \citenamefont {Principi}, \citenamefont {Torre},
  \citenamefont {Ben~Shalom}, \citenamefont {Taniguchi}, \citenamefont
  {Watanabe}, \citenamefont {Grigorieva}, \citenamefont {Polini}, \citenamefont
  {Geim},\ and\ \citenamefont {Bandurin}}]{Berdyugineaau0685}%
  \BibitemOpen
  \bibfield  {author} {\bibinfo {author} {\bibfnamefont {A.~I.}\ \bibnamefont
  {Berdyugin}}, \bibinfo {author} {\bibfnamefont {S.~G.}\ \bibnamefont {Xu}},
  \bibinfo {author} {\bibfnamefont {F.~M.~D.}\ \bibnamefont {Pellegrino}},
  \bibinfo {author} {\bibfnamefont {R.}~\bibnamefont {Krishna~Kumar}}, \bibinfo
  {author} {\bibfnamefont {A.}~\bibnamefont {Principi}}, \bibinfo {author}
  {\bibfnamefont {I.}~\bibnamefont {Torre}}, \bibinfo {author} {\bibfnamefont
  {M.}~\bibnamefont {Ben~Shalom}}, \bibinfo {author} {\bibfnamefont
  {T.}~\bibnamefont {Taniguchi}}, \bibinfo {author} {\bibfnamefont
  {K.}~\bibnamefont {Watanabe}}, \bibinfo {author} {\bibfnamefont {I.~V.}\
  \bibnamefont {Grigorieva}}, \bibinfo {author} {\bibfnamefont
  {M.}~\bibnamefont {Polini}}, \bibinfo {author} {\bibfnamefont {A.~K.}\
  \bibnamefont {Geim}}, \ and\ \bibinfo {author} {\bibfnamefont {D.~A.}\
  \bibnamefont {Bandurin}},\ }\href@noop {} {\bibfield  {journal} {\bibinfo
  {journal} {Science}\ }\textbf {\bibinfo {volume} {364}},\ \bibinfo {pages}
  {162} (\bibinfo {year} {2019})}\BibitemShut {NoStop}%
\bibitem [{\citenamefont {Delacretaz}\ and\ \citenamefont
  {Gromov}(2017)}]{Delacretaz:2017yia}%
  \BibitemOpen
  \bibfield  {author} {\bibinfo {author} {\bibfnamefont {L.~V.}\ \bibnamefont
  {Delacretaz}}\ and\ \bibinfo {author} {\bibfnamefont {A.}~\bibnamefont
  {Gromov}},\ }\href {\doibase 10.1103/PhysRevLett.120.079901,
  10.1103/PhysRevLett.119.226602} {\bibfield  {journal} {\bibinfo  {journal}
  {Phys. Rev. Lett.}\ }\textbf {\bibinfo {volume} {119}},\ \bibinfo {pages}
  {226602} (\bibinfo {year} {2017})},\ \bibinfo {note} {phys. Rev. Lett.
  \textbf{120}, 079901 (2018)}\BibitemShut {NoStop}%
\bibitem [{\citenamefont {{Pellegrino}}\ \emph {et~al.}(2017)\citenamefont
  {{Pellegrino}}, \citenamefont {{Torre}},\ and\ \citenamefont
  {{Polini}}}]{Pellegrino:2017fd}%
  \BibitemOpen
  \bibfield  {author} {\bibinfo {author} {\bibfnamefont {F.~M.~D.}\
  \bibnamefont {{Pellegrino}}}, \bibinfo {author} {\bibfnamefont
  {I.}~\bibnamefont {{Torre}}}, \ and\ \bibinfo {author} {\bibfnamefont
  {M.}~\bibnamefont {{Polini}}},\ }\href {\doibase 10.1103/PhysRevB.96.195401}
  {\bibfield  {journal} {\bibinfo  {journal} {Phys. Rev. B}\ }\textbf {\bibinfo
  {volume} {96}},\ \bibinfo {eid} {195401} (\bibinfo {year}
  {2017})}\BibitemShut {NoStop}%
\bibitem [{\citenamefont {{Holder}}\ \emph {et~al.}(2019)\citenamefont
  {{Holder}}, \citenamefont {{Queiroz}},\ and\ \citenamefont
  {{Stern}}}]{Stern2019}%
  \BibitemOpen
  \bibfield  {author} {\bibinfo {author} {\bibfnamefont {T.}~\bibnamefont
  {{Holder}}}, \bibinfo {author} {\bibfnamefont {R.}~\bibnamefont {{Queiroz}}},
  \ and\ \bibinfo {author} {\bibfnamefont {A.}~\bibnamefont {{Stern}}},\
  }\href@noop {} {\bibfield  {journal} {\bibinfo  {journal} {arXiv:1903.05541}\
  } (\bibinfo {year} {2019})}\BibitemShut {NoStop}%
\bibitem [{\citenamefont {{Sulpizio}}\ \emph {et~al.}()\citenamefont
  {{Sulpizio}}, \citenamefont {{Ella}}, \citenamefont {{Rozen}}, \citenamefont
  {{Birkbeck}}, \citenamefont {{Perello}}, \citenamefont {{Dutta}},
  \citenamefont {{Ben-Shalom}}, \citenamefont {{Taniguchi}}, \citenamefont
  {{Watanabe}}, \citenamefont {{Holder}}, \citenamefont {{Queiroz}},
  \citenamefont {{Stern}}, \citenamefont {{Scaffidi}}, \citenamefont {{Geim}},\
  and\ \citenamefont {{Ilani}}}]{2019arXiv190511662S}%
  \BibitemOpen
  \bibfield  {author} {\bibinfo {author} {\bibfnamefont {J.~A.}\ \bibnamefont
  {{Sulpizio}}}, \bibinfo {author} {\bibfnamefont {L.}~\bibnamefont {{Ella}}},
  \bibinfo {author} {\bibfnamefont {A.}~\bibnamefont {{Rozen}}}, \bibinfo
  {author} {\bibfnamefont {J.}~\bibnamefont {{Birkbeck}}}, \bibinfo {author}
  {\bibfnamefont {D.~J.}\ \bibnamefont {{Perello}}}, \bibinfo {author}
  {\bibfnamefont {D.}~\bibnamefont {{Dutta}}}, \bibinfo {author} {\bibfnamefont
  {M.}~\bibnamefont {{Ben-Shalom}}}, \bibinfo {author} {\bibfnamefont
  {T.}~\bibnamefont {{Taniguchi}}}, \bibinfo {author} {\bibfnamefont
  {K.}~\bibnamefont {{Watanabe}}}, \bibinfo {author} {\bibfnamefont
  {T.}~\bibnamefont {{Holder}}}, \bibinfo {author} {\bibfnamefont
  {R.}~\bibnamefont {{Queiroz}}}, \bibinfo {author} {\bibfnamefont
  {A.}~\bibnamefont {{Stern}}}, \bibinfo {author} {\bibfnamefont
  {T.}~\bibnamefont {{Scaffidi}}}, \bibinfo {author} {\bibfnamefont {A.~K.}\
  \bibnamefont {{Geim}}}, \ and\ \bibinfo {author} {\bibfnamefont
  {S.}~\bibnamefont {{Ilani}}},\ }\href@noop {} {\ }\Eprint
  {http://arxiv.org/abs/1905.11662} {arXiv:1905.11662} \BibitemShut {NoStop}%
\bibitem [{Note1()}]{Note1}%
  \BibitemOpen
  \bibinfo {note} {In what follows, we assume that the screening length in our
  material is much shorter than the width of our channel. Hence, the electric
  field sourced by the redistribution of charge carriers in our fluid can be
  neglected.}\BibitemShut {Stop}%
\bibitem [{\citenamefont {Erdmenger}\ \emph {et~al.}(2018)\citenamefont
  {Erdmenger}, \citenamefont {Matthaiakakis}, \citenamefont {Meyer},\ and\
  \citenamefont {Fern{\'a}ndez}}]{erdmenger2018strongly}%
  \BibitemOpen
  \bibfield  {author} {\bibinfo {author} {\bibfnamefont {J.}~\bibnamefont
  {Erdmenger}}, \bibinfo {author} {\bibfnamefont {I.}~\bibnamefont
  {Matthaiakakis}}, \bibinfo {author} {\bibfnamefont {R.}~\bibnamefont
  {Meyer}}, \ and\ \bibinfo {author} {\bibfnamefont {D.~R.}\ \bibnamefont
  {Fern{\'a}ndez}},\ }\href@noop {} {\bibfield  {journal} {\bibinfo  {journal}
  {Phys. Rev. B}\ }\textbf {\bibinfo {volume} {98}},\ \bibinfo {pages} {195143}
  (\bibinfo {year} {2018})}\BibitemShut {NoStop}%
\bibitem [{Note2()}]{Note2}%
  \BibitemOpen
  \bibinfo {note} {The charge conservation equation $\partial _t \rho + \nabla
  \cdot (\rho \protect \mathbf {v})=0$ is trivially satisfied due to our
  assumption of a steady and translational invariant flow $\protect \mathbf
  {v}=v_\protect \mathrm {x}(y) \protect \mathbf {e}_\protect \mathrm
  {x}$.}\BibitemShut {Stop}%
\bibitem [{Note3()}]{Note3}%
  \BibitemOpen
  \bibinfo {note} {In general, one could also allow for a temperature gradient
  $T \protect \tmspace -\thinmuskip {.1667em} \to \protect \tmspace
  -\thinmuskip {.1667em} T \protect \tmspace -\thinmuskip {.1667em} + \protect
  \tmspace -\thinmuskip {.1667em} \Delta T$. However, we show in Ref.~\cite
  {supp} that this contribution is negligibly small for Fermi liquids. In
  addition, since we restrict ourselves to $T = \protect \mathcal
  {O}(1{\protect \rm K})$, phonons are negligible in our approach \cite
  {Molenkamp:1994ii,Molenkamp:1994kb}.}\BibitemShut {Stop}%
\bibitem [{\citenamefont {Landau}\ and\ \citenamefont
  {Lifshitz}(1987)}]{Landau:1987}%
  \BibitemOpen
  \bibfield  {author} {\bibinfo {author} {\bibfnamefont {L.}~\bibnamefont
  {Landau}}\ and\ \bibinfo {author} {\bibfnamefont {E.}~\bibnamefont
  {Lifshitz}},\ }\href@noop {} {\emph {\bibinfo {title} {Fluid Mechanics}}},\
  \bibinfo {edition} {2nd}\ ed.,\ Course of Theoretical Physics S\ (\bibinfo
  {publisher} {Butterworth-Heinemann},\ \bibinfo {year} {1987})\BibitemShut
  {NoStop}%
\bibitem [{\citenamefont {Romatschke}(2010)}]{Romatschke:2010jf}%
  \BibitemOpen
  \bibfield  {author} {\bibinfo {author} {\bibfnamefont {P.}~\bibnamefont
  {Romatschke}},\ }\href@noop {} {\bibfield  {journal} {\bibinfo  {journal}
  {Int. J. Mod. Phys. E}\ }\textbf {\bibinfo {volume} {19}},\ \bibinfo {pages}
  {1} (\bibinfo {year} {2010})}\BibitemShut {NoStop}%
\bibitem [{\citenamefont {Luciano~Rezzolla}(2013)}]{Rezzolla:2013}%
  \BibitemOpen
  \bibfield  {author} {\bibinfo {author} {\bibfnamefont {O.~Z.}\ \bibnamefont
  {Luciano~Rezzolla}},\ }\href@noop {} {\emph {\bibinfo {title} {Relativistic
  Hydrodynamics}}}\ (\bibinfo  {publisher} {Oxford University Press},\ \bibinfo
  {year} {2013})\BibitemShut {NoStop}%
\bibitem [{sup()}]{supp}%
  \BibitemOpen
  \href@noop {} {\bibinfo  {journal} {See supplemental material for further
  details}\ }\BibitemShut {NoStop}%
\bibitem [{Note4()}]{Note4}%
  \BibitemOpen
\bibfield  {journal} {  }\bibinfo {note} {Let us emphasize that we took into
  account the term $\propto \protect \tmspace -\thinmuskip {.1667em} \Delta \mu
  E_\protect \mathrm {x}$, even though it is higher order in $E_\protect
  \mathrm {x}$ and therefore does not affect our linear response results. We
  did so to explicitly see that the Lorentz and the Hall-viscosity induced
  force influences the velocity profile in the expected way (cf.
  Ref.~[29]).}\BibitemShut {Stop}%
\bibitem [{\citenamefont {Scaffidi}\ \emph {et~al.}(2017)\citenamefont
  {Scaffidi}, \citenamefont {Nandi}, \citenamefont {Schmidt}, \citenamefont
  {Mackenzie},\ and\ \citenamefont {Moore}}]{scaffidi2017hydrodynamic}%
  \BibitemOpen
  \bibfield  {author} {\bibinfo {author} {\bibfnamefont {T.}~\bibnamefont
  {Scaffidi}}, \bibinfo {author} {\bibfnamefont {N.}~\bibnamefont {Nandi}},
  \bibinfo {author} {\bibfnamefont {B.}~\bibnamefont {Schmidt}}, \bibinfo
  {author} {\bibfnamefont {A.~P.}\ \bibnamefont {Mackenzie}}, \ and\ \bibinfo
  {author} {\bibfnamefont {J.~E.}\ \bibnamefont {Moore}},\ }\href@noop {}
  {\bibfield  {journal} {\bibinfo  {journal} {Phys. Rev. Lett.}\ }\textbf
  {\bibinfo {volume} {118}},\ \bibinfo {pages} {226601} (\bibinfo {year}
  {2017})}\BibitemShut {NoStop}%
\bibitem [{Note5()}]{Note5}%
  \BibitemOpen
  \bibinfo {note} {Notice that we do not use standard Robin type boundary
  conditions to define the slip length in our system \cite {Kiselev18}. This
  relies on the fact that our set of boundary conditions naturally describes
  the case of a finite drift velocity.In our supplemental material we prove
  that in our channel geometry, Robin-type boundary conditions are equivalent
  with our choice of boundary conditions, Eq.\protect \textup {\hbox
  {\mathsurround \z@ \protect \normalfont (\ignorespaces \ref {eq:bsc}\unskip
  \@@italiccorr )}}.}\BibitemShut {Stop}%
\bibitem [{Note6()}]{Note6}%
  \BibitemOpen
  \bibinfo {note} {In this limit, according to Eq.~\protect \textup {\hbox
  {\mathsurround \z@ \protect \normalfont (\ignorespaces \ref
  {eq:HallVoltagesCase3}\unskip \@@italiccorr )}}, $\Delta V_{\eta _\protect
  \mathrm {H}}$ becomes insensitive to the concrete form of the boundary
  conditions.}\BibitemShut {Stop}%
\bibitem [{\citenamefont {{Moessner}}\ \emph {et~al.}(2019)\citenamefont
  {{Moessner}}, \citenamefont {{Morales--Dur{\'a}n}}, \citenamefont
  {{Sur{\'o}wka}},\ and\ \citenamefont {{Witkowski}}}]{Moessner19}%
  \BibitemOpen
  \bibfield  {author} {\bibinfo {author} {\bibfnamefont {R.}~\bibnamefont
  {{Moessner}}}, \bibinfo {author} {\bibfnamefont {N.}~\bibnamefont
  {{Morales--Dur{\'a}n}}}, \bibinfo {author} {\bibfnamefont {P.}~\bibnamefont
  {{Sur{\'o}wka}}}, \ and\ \bibinfo {author} {\bibfnamefont {P.}~\bibnamefont
  {{Witkowski}}},\ }\href@noop {} {\bibfield  {journal} {\bibinfo  {journal}
  {arXiv:1903.08037}\ } (\bibinfo {year} {2019})}\BibitemShut {NoStop}%
\bibitem [{\citenamefont {Giuliani}\ and\ \citenamefont
  {Quinn}(1982)}]{PhysRevB.26.4421}%
  \BibitemOpen
  \bibfield  {author} {\bibinfo {author} {\bibfnamefont {G.~F.}\ \bibnamefont
  {Giuliani}}\ and\ \bibinfo {author} {\bibfnamefont {J.~J.}\ \bibnamefont
  {Quinn}},\ }\href {\doibase 10.1103/PhysRevB.26.4421} {\bibfield  {journal}
  {\bibinfo  {journal} {Phys. Rev. B}\ }\textbf {\bibinfo {volume} {26}},\
  \bibinfo {pages} {4421} (\bibinfo {year} {1982})}\BibitemShut {NoStop}%
\bibitem [{\citenamefont {Czycholl}(2016)}]{bookref}%
  \BibitemOpen
  \bibfield  {author} {\bibinfo {author} {\bibfnamefont {G.}~\bibnamefont
  {Czycholl}},\ }\href@noop {} {\emph {\bibinfo {title} {Theoretische
  Festk\"orperphysik Band 1}}}\ (\bibinfo  {publisher} {Springer Berlin
  Heidelberg},\ \bibinfo {year} {2016})\BibitemShut {NoStop}%
\bibitem [{Note10()}]{Note10}%
  \BibitemOpen
  \bibinfo {note} {Note that the system lies within the range of validity of
  the hydrodynamic regime, even though $l_{\protect \rm ee}\lesssim W $ \cite
  {2019arXiv190511662S}.}\BibitemShut {Stop}%
\bibitem [{Note7()}]{Note7}%
  \BibitemOpen
  \bibinfo {note} {Private communication with Prof. Laurens Molenkamp,
  University of Wuerzburg}\BibitemShut {NoStop}%
\bibitem [{\citenamefont {B\"ottcher}\ \emph {et~al.}(2019)\citenamefont
  {B\"ottcher}, \citenamefont {Tutschku}, \citenamefont {Molenkamp},\ and\
  \citenamefont {Hankiewicz}}]{Tutschku19}%
  \BibitemOpen
  \bibfield  {author} {\bibinfo {author} {\bibfnamefont {J.}~\bibnamefont
  {B\"ottcher}}, \bibinfo {author} {\bibfnamefont {C.}~\bibnamefont
  {Tutschku}}, \bibinfo {author} {\bibfnamefont {L.~W.}\ \bibnamefont
  {Molenkamp}}, \ and\ \bibinfo {author} {\bibfnamefont {E.~M.}\ \bibnamefont
  {Hankiewicz}},\ }\href@noop {} {\bibfield  {journal} {\bibinfo  {journal}
  {arXiv:1901.05425}\ } (\bibinfo {year} {2019})}\BibitemShut {NoStop}%
\bibitem [{\citenamefont {Hughes}\ \emph {et~al.}(2013)\citenamefont {Hughes},
  \citenamefont {Leigh},\ and\ \citenamefont {Parrikar}}]{Hughes12}%
  \BibitemOpen
  \bibfield  {author} {\bibinfo {author} {\bibfnamefont {T.~L.}\ \bibnamefont
  {Hughes}}, \bibinfo {author} {\bibfnamefont {R.~G.}\ \bibnamefont {Leigh}}, \
  and\ \bibinfo {author} {\bibfnamefont {O.}~\bibnamefont {Parrikar}},\
  }\href@noop {} {\bibfield  {journal} {\bibinfo  {journal} {Phys. Rev. D}\
  }\textbf {\bibinfo {volume} {88}},\ \bibinfo {pages} {025040} (\bibinfo
  {year} {2013})}\BibitemShut {NoStop}%
\bibitem [{\citenamefont {Kiselev}\ and\ \citenamefont
  {Schmalian}(2019)}]{Kiselev18}%
  \BibitemOpen
  \bibfield  {author} {\bibinfo {author} {\bibfnamefont {E.~I.}\ \bibnamefont
  {Kiselev}}\ and\ \bibinfo {author} {\bibfnamefont {J.}~\bibnamefont
  {Schmalian}},\ }\href {\doibase 10.1103/PhysRevB.99.035430} {\bibfield
  {journal} {\bibinfo  {journal} {Phys. Rev. B}\ }\textbf {\bibinfo {volume}
  {99}},\ \bibinfo {pages} {035430} (\bibinfo {year} {2019})}\BibitemShut
  {NoStop}%
\bibitem [{\citenamefont {{Novikov}}(2006)}]{Novikov06}%
  \BibitemOpen
  \bibfield  {author} {\bibinfo {author} {\bibfnamefont {D.~S.}\ \bibnamefont
  {{Novikov}}},\ }\href@noop {} {\bibfield  {journal} {\bibinfo  {journal}
  {arXiv:0603184}\ } (\bibinfo {year} {2006})}\BibitemShut {NoStop}%
\bibitem [{\citenamefont {{Qian}}\ and\ \citenamefont
  {{Vignale}}(2005)}]{Qian05}%
  \BibitemOpen
  \bibfield  {author} {\bibinfo {author} {\bibfnamefont {Z.}~\bibnamefont
  {{Qian}}}\ and\ \bibinfo {author} {\bibfnamefont {G.}~\bibnamefont
  {{Vignale}}},\ }\href@noop {} {\bibfield  {journal} {\bibinfo  {journal}
  {Phys. Rev. B}\ }\textbf {\bibinfo {volume} {71}},\ \bibinfo {eid} {075112}
  (\bibinfo {year} {2005})}\BibitemShut {NoStop}%
\end{thebibliography}%


\appendix

\renewcommand{\thefigure}{S\arabic{figure}}
\renewcommand{\theequation}{S\arabic{equation}}
\setcounter{equation}{0}

\section*{Supplemental material}

\title{Supplemental material: ``Functional dependence of the Hall viscosity-induced transverse voltage in two-dimensional Fermi liquids"}

\author{Ioannis~Matthaiakakis}
\thanks{These three authors contributed equally to this work and are listed alphabetically.}
\affiliation{Institute for Theoretical Physics and Astrophysics and W\"urzburg-Dresden Cluster of Excellence ct.qmat,
Julius-Maximilians-Universit\"at W\"urzburg, 97074 W\"urzburg, Germany}

\author{David~Rodr\'iguez~Fern\'andez}
\thanks{These three authors contributed equally to this work and are listed alphabetically.}
\affiliation{Institute for Theoretical Physics and Astrophysics and W\"urzburg-Dresden Cluster of Excellence ct.qmat,
Julius-Maximilians-Universit\"at W\"urzburg, 97074 W\"urzburg, Germany}

\author{\mbox{Christian~Tutschku}}
\thanks{These three authors contributed equally to this work and are listed alphabetically.}
\affiliation{Institute for Theoretical Physics and Astrophysics and W\"urzburg-Dresden Cluster of Excellence ct.qmat,
Julius-Maximilians-Universit\"at W\"urzburg, 97074 W\"urzburg, Germany}

\author{Ewelina~M. Hankiewicz}
\affiliation{Institute for Theoretical Physics and Astrophysics and W\"urzburg-Dresden Cluster of Excellence ct.qmat,
Julius-Maximilians-Universit\"at W\"urzburg, 97074 W\"urzburg, Germany}

\author{Johanna~Erdmenger}
\affiliation{Institute for Theoretical Physics and Astrophysics and W\"urzburg-Dresden Cluster of Excellence ct.qmat,
Julius-Maximilians-Universit\"at W\"urzburg, 97074 W\"urzburg, Germany}

\author{Ren\'e~Meyer}
\affiliation{Institute for Theoretical Physics and Astrophysics and W\"urzburg-Dresden Cluster of Excellence ct.qmat,
Julius-Maximilians-Universit\"at W\"urzburg, 97074 W\"urzburg, Germany}

\maketitle

\textit{Equations of motion.} In what follows, we derive the equations of motion for a charged two-dimensional Fermi liquid confined to a channel geometry, in the framework of linear response theory. In particular, we study fluid dynamics in presence of a DC electric field $\mathbf{E}$ and an out-of-plane magnetic field $\mathbf{B}$. For such a fluid, the equations of motion are given by charge continuity  as well as by momentum conservation (Navier-Stokes):%
\begin{align}
\label{eq:HydroEq} 
\!  \left(\partial_t + \mathbf{v}\cdot\nabla\right)\rho  &=  -\rho \nabla \cdot \mathbf{v} ,
\\
\! m_{\rm eff}\rho \left( \partial_t + \mathbf{v}\cdot\nabla \right)\mathbf{v}  &=  -\nabla p  + \eta \nabla^2 \mathbf{v} +\eta_{\rm H} \nabla^2(\mathbf{v}\times \mathbf{e}_{\rm z}) \nonumber
\\
&~~~~ + {\rm e}\rho (\mathbf{E} + \mathbf{v}\times \mathbf{B}) - \dfrac{\rho_0 v_{\rm F}m_{\rm eff} }{ l_{\rm imp}}\mathbf{v}  . \label{eq:NSEq}
\end{align}
\noindent
In our analysis, we consider a steady, hydrodynamic flow of electrons, which is translationally invariant along the $\mathbf{e}_{\rm x}$-direction. Moreover, we assume a vanishing current in the $\mathbf{e}_{\rm y}$-direction, implying that the velocity profile takes the form $\mathbf{v} = v_{\rm x}(y) \mathbf{e}_{\rm x}$. To obtain an inhomogeneous, non-trivial velocity profile, our system needs to deviate from global thermal equilibrium. However, in order to be able to linearize Eqs.~\eqref{eq:HydroEq} and \eqref{eq:NSEq}, we assume that  variations of the chemical potential and temperature are small compared to their global equilibrium values
\begin{align} \label{globaleq}
\mu (y) & = \mu_0 + \Delta \mu (y) \quad \text{with} \ \ \; \;  \Delta \mu (y)\ll \mu_0 \ \,  , 
\\
T(y) & = T_0 + \Delta T(y) \quad \text{with} \quad \Delta T(y)\ll T_0 \ . \nonumber
\end{align}
For typical Fermi liquids, such as GaAs, $\mu_0 = {\cal O}(50 {\rm ~meV})$ whereas $T_0 = {\cal O}(1{\rm~ K})$. Note, that due to time and translational  invariance in  $\mathbf{e}_{\rm x}$-direction, $\Delta\mu$ and $\Delta T$ solely vary in 
the $\mathbf{e}_{\rm y}$-direction. In terms of $\mu(y)$ and $T(y)$,  pressure and  density fluctuations in our system are given by
\begin{align} \label{densiytypressureinterms}
&\rho (y) = \rho_0 + \dfrac{\partial \rho_0 }{ \partial \mu_0} \Delta\mu + \dfrac{\partial \rho_0 }{ \partial T_0} \Delta T \, ,   
\\
&p(y) = p_0 + \dfrac{\partial p_0}{\partial \mu_0}\Delta \mu + \dfrac{\partial p_0}{\partial T} \Delta T  = p_0 + \rho_0\Delta \mu + s_0 \Delta T \, , \nonumber
\end{align}
where $s_0$ is the equilibrium entropy of our fluid.
Under the assumption of a steady and unidirectional fluid flow with $v_\mathrm{y}=0$, Eq.~\eqref{eq:HydroEq} reduces to $\partial_{\rm x}\rho =0$.
This enforces a constant density along the fluid flow, whereas it allows for density fluctuations in 
$\mathrm{\mathbf{e}}_\mathrm{y}$-direction.
Explicitly, that is compatible with our incompressibility condition \mbox{$\nabla\cdot  \mathbf{v}\!=\!0$}. Together, Eqs.~\eqref{eq:HydroEq}-\eqref{densiytypressureinterms} dictate the dynamics of $v_{\rm x}(y)$, $\Delta \mu (y)$ and $\Delta T(y)$. 
For our system the gradient of temperature is negligible compared to the gradient of the chemical potential. To see this explicitly, we write
\begin{align} \label{eq:gradp}
\partial_y p = \rho_0\mu_0 \partial_y\left(\dfrac{\Delta \mu }{ \mu_0 } \right) + s_0 T_0 \partial_y\left(\dfrac{\Delta T }{ T_0} \right) \ .
\end{align}
Due to the assumptions in Eq.~\eqref{globaleq}, the 
dimensionless gradients $\partial_y(\Delta \mu/\mu_0)$ and $\partial_y(\Delta T/ T_0)$ are of the same order of magnitude. As a consequence, the relative strength of the chemical potential contribution to the temperature contribution in Eq.~\eqref{eq:gradp} is given by ${\cal U} = \rho_0 \mu_0 / s_0 T_0$.
If ${\cal U}\gg 1$, the chemical potential gradient dominates over temperature fluctuations and vice versa. For typical Fermi liquids, such as GaAs, we find ${\cal U} \simeq 10^{16}$. Hence, in those samples the gradient of temperature is clearly negligible. Thus, we assume $T(y) = T_0$ and define
\begin{align}
&\rho (y) = \rho_0 + \dfrac{\partial \rho_0 }{ \partial \mu_0} \Delta \mu(y) \ 
\wedge \  p(y) = p_0 + \rho_0\Delta \mu(y)\ . 
\end{align}
Substituting this into Eq.~\eqref{eq:NSEq}, leads to
\begin{align} \label{eq:NPNS1'}
\eta \partial^2_{\rm y} v_{\rm x}(y) \! & = \!  \left( \rho_0 + \dfrac{\partial \rho_0 }{ \partial \mu} \Delta \mu\right) \! \! \left({\rm e}E_{\rm x} + \dfrac{v_{\rm F}m_{\rm eff} }{ l_{\rm imp}}v_{\rm x}(y)\right)  , 
\\
\partial_{\rm y} p  & =  \left[{\rm e}B\left(\rho_0 + \dfrac{\partial \rho_0 }{ \partial \mu} \Delta \mu\right) -\eta_{\rm H}\partial^2_y\right]v_{\rm x}(y) \ .\label{eq:NPNS2'}
\end{align}
Let us emphasize that we did not include the electric field sourced by our inhomogeneous density distribution $\rho(y)$ in Eqs.~\eqref{eq:NPNS1'} and \eqref{eq:NPNS2'}. 
This is justified by the assumption that the screening length of our material is much shorter than the width of our channel. Hence, the corresponding electric field 
is expected to be much smaller than $E_{\rm x}$, and therefore can be neglected. 

So far, Eqs.~\eqref{eq:NPNS1'} and \eqref{eq:NPNS2'} still deviate from Eqs.~(2) and (3) of our main text, due to the presence of additional terms $\propto \! \Delta \mu(y) v_{\rm x}(y)$. 
These terms induce non-linear corrections to our physical observables in terms of the electric field $E_{\rm x}$. In the framework of linear response theory, we are therefore allowed to drop them without loss of generality. In particular, this leads to Eqs.~(2) and (3) of our main text:
\begin{align}
\label{eq:NPNS1}
&\eta \partial^2_{\rm y} v_{\rm x}(y) = \left( \rho_0 + \dfrac{\partial \rho_0}{ \partial \mu} \Delta \mu\right){\rm e}E_{\rm x} + \dfrac{\rho_0v_{\rm F}m_{\rm eff} }{ l_{\rm imp}} v_{\rm x}(y)  , 
\\
&\partial_{\rm y} p = \rho_0\partial_y \Delta \mu = \left({\rm e}B\rho_0 -\eta_{\rm H}\partial^2_y\right)v_{\rm x}(y) \ . \label{eq:NPNS2}
\end{align}
Notice, that we took into account the term $\propto \! \Delta \mu(y) E_{\rm x}$, even though it also scales non-linearly in $E_\mathrm{x}$ and, therefore, does not affect our phyiscal observables. However, we kept this term because we want to explicitly check that the Lorentz and the Hall-viscosity induced force influences the velocity profile in the expected way. Namely, we expect $v_\mathrm{x}(y)$ to deviate from its symmetric Poiseuille form, corresponding to a fluid flow which concentrates towards one of the sides of the channel, depending on the sign of the magnetic field. The simplest way to see this deviation is to take into account the $\Delta \mu E_\mathrm{x}$ term since it only couples  to the first order correction of the velocity profile and decouples from the equations for $v^{(0)}_\mathrm{ x}(y)$.

For our setup, we supplement Eqs.~\eqref{eq:NPNS1} and \eqref{eq:NPNS2} with general boundary conditions of the form
\begin{align}
\label{eq:bcs} 
v_{\rm x}(-l_{\rm s}) =  v_{\rm x}(W + l_{\rm s}) & =  0    \ ,  \\   \nonumber
\Delta  \mu \big \vert_{y=0}    + \ \Delta \mu \big \vert_{y=W} \ \ & = 0 \ . 
\end{align}
Below, in section \emph{Robin Boundary Conditions}, we prove that this choice is equivalent to standard Robin type boundary conditions, as defined in Ref.~\cite{Kiselev18}. In what follows, we provide an explicit solution of the Navier-Stokes Eqs.~\eqref{eq:NPNS1}-\eqref{eq:NPNS2} under the boundary condition~\eqref{eq:bcs} for weak and for strong magnetic fields, since in this regime the physical properties of the analytic solutions become most apparent.\\

\textit{Weak magnetic field regime.}  
The weak magnetic field limit $r_\mathrm{c} \gg l_{\rm G}$ 
allows for the expansion of the velocity profile and hence of the Navier-Stokes
equations in powers of $B$. Technically, this expansion is achieved by introducing the dimensionless  parameter $\epsilon \ll 1$, satisfying
\begin{align}
\! B \rightarrow \epsilon B , \, \eta_{\rm H} \rightarrow \epsilon \eta_{\rm H}, \, \Delta \mu \rightarrow \epsilon\Delta \mu  \, , v_{\rm x} \! = \! v^0_{\rm x} + \epsilon v^1_{\rm x}.
\end{align}
In particular, this assumes that to first order the system  responds linearly in terms of the magnetic field. 
With this ansatz the Navier-Stokes
Eqs.~\eqref{eq:NPNS1}-\eqref{eq:NPNS2} reduce to 
\begin{align}
\label{eq:WeakB NSE1}
\eta \partial^2_{\rm y}v^0_{\rm x}(y) -\frac{ m_{\rm eff}v_{\rm F}\rho_0}{l_{\rm imp} }v^0_{ \rm x}(y) & ={\rm e}\rho_0 E_\mathrm{x} \, , 
\\
\left({\rm e}B \rho_0 - \eta_H \partial^2_{\rm y}   \right)v^0_{\rm x}(y)  & = \rho_0 \partial_{\rm y}\Delta \mu(y) \, , \label{eq:WeakB NSE2}
\\
\eta \partial^2_{\rm y}v^1_{\rm x}(y) -\frac{ m_{\rm eff}v_{\rm F}\rho_0}{l_{\rm imp}  } v^1_{\rm x}(y) & = {\rm e} E_{\rm x} \dfrac{\partial \rho_0 }{ \partial\mu} \Delta \mu(y) \, . \label{eq:WeakB NSE3}
\end{align}
To find an explicit solution of this set of equations we first 
determine  $v^0_{\rm x}(y)$ by solving  Eq.~\eqref{eq:WeakB NSE1}, which is the zero-field Poiseuille flow equation in the presence of impurities. Once we have obtained $v^0_{\rm x}(y)$, we calculate $\Delta \mu(y)$ by 
integrating Eq.~\eqref{eq:WeakB NSE2}.
Substituting this quantity in Eq.~\eqref{eq:WeakB NSE3} finally allows us to derive the first order velocity correction $v^1_{\rm x}(y)$. 
Explicitly, we find
\begin{align}
\label{eq:WeakB vx0}
\! \! \! \!  \! \! v^0_{\rm x}(y) \! & = \! - \dfrac{{\rm e} E_{\rm x}l_{\rm imp} }{ m_{\rm eff}v_{\rm F}} \! \! \left( \! A_1 \! \cosh \! \left[   yl_{\rm G}^{-1}  \right] \! \! + \! A_2 \! \sinh \! \left[   yl_{\rm G}^{-1}  \right] \! \! + \! 1   \right) 
\\
\label{eq:WeakB mu}
\! \!  \! \! \! \!  \Delta \mu(y)  \! & = \!  \frac{\mathrm{e} \,l_{\rm imp}\,   E_\mathrm{x}}{m_{\rm eff} v_{\rm F} } \, \Big[  l_{\rm G} \,  \Big( \frac{m_{\rm eff} v_{\rm F} \eta_{\rm H}}{ \eta l_{\rm imp} }-{\rm e}B \Big)  \\
& \times \left( A_1 \! \sinh \! \left[y l_{\rm G}^{-1}\right] + A_2 \! \cosh \! \left[y l_{\rm G}^{-1} \right] \right)  \! -\!  {\rm e}By \Big] + \Gamma, \nonumber
\end{align}
where for clarity we defined 
\begin{align}
\label{eq:vx0 constants} 
A_1 & = -\cosh \left[\frac{W}{2 l_{\rm G}}\right] {\rm sech}\left[\frac{2 l_{\rm s}+W}{2 l_{\rm G}}\right], 
\\
\label{eq:vx0 constants'}
A_2 & = \ \; \; \, \sinh \left[\frac{W}{2 l_{\rm G}}\right] {\rm sech}\left[\frac{2 l_\mathrm{s}+W}{2 l_{\rm G}}\right].
\\
\label{eq:Gamma}
\! \Gamma &=-\frac{\mathrm{e} \, l_{\rm imp}   E_\mathrm{x}}{2m_{\rm eff} v_{\rm F}} \, \Big[  l_{\rm G} \,  \Big( \frac{m_{\rm eff} v_{\rm F} \eta_{\rm H}}{ \eta l_{\rm imp} }-{\rm e}B \Big) \\
& \times \Big( A_1 \!
\sinh \! \left[ W l_{\rm G}^{-1}\right] + A_2 \! \left[\cosh \! \left[W l_{\rm G}^{-1} \Big]+1\right] \right)  \! -\! {\rm e}BW \Big] \ \nonumber . 
\end{align} 
Before we explicitly present our result for $v^1_{\rm x}(y)$, we 
want to emphasize that Eq.~\eqref{eq:WeakB mu}  predicts 
the total Hall voltage $\Delta  V_{\rm tot} = -[\Delta \mu(W) \! - \! \Delta \mu(0)]/ \mathrm{e}$, measured across the sample. To derive Eq.~(8) of the main text, we need to expand the hyperbolic functions in Eqs.~\eqref{eq:vx0 constants}-\eqref{eq:Gamma} up to third order in $W/l_{\rm G}$. As a result, we find 
\begin{align} \label{eq:HallVoltagesCase32}
\! \Delta 	V_{\eta_{\rm H}} \! \!  &= \!  \frac{\eta_\mathrm{H} E_\mathrm{x} }{\eta}   \! \! \left[ W \! \! - \!  
  { 1 \over  12    l_\mathrm{ G}^2 } \!
 \left( W^3 \! \! + \! 6 l_{\rm s} W^2 \! \! + \! 6 l_{\rm s}^2 W\right) \!  \right]  \\
\Delta 	V_{B} \! &= \!   - \frac{ \mathrm{sgn}(B)  E_\mathrm{x}}{3 r_\mathrm{c} l_\mathrm{ee} } \left( W^3 \! + \! 6 l_{\rm s} W^2 \! + \! 6 l_{\rm s}^2 W\right) \ . \label{eq:lorentz}
\end{align}
In Eq.~(10) of the main text, we have illustrated the density dependence of these voltage contributions,
\begin{align} \label{eq:densitydep2}
	\Delta  V_{\eta_\mathrm{H}} & = f_1 \, \rho_0 + f_2(n_\mathrm{imp})  \quad  \wedge \quad
\Delta V_B = f_3 \, \rho_0^{-2}  \ .
\end{align}  
However, so far we did not specify the functions $f_{1,2,3}$. This will be  the purpose of the following paragraph. The linear, impurity independent scaling of $\Delta V_{ \eta_{\rm H}}$ is defined as the $l_\mathrm{imp} \! \rightarrow \! \infty$ limit of Eq.~\eqref{eq:HallVoltagesCase32}. The density dependence of this term is given by (cf. Eq.~(4) of our main text)
\begin{align}
\Delta V_{\eta_{\rm H}}^0 &= \dfrac{\rm \eta_{\rm H} }{ \eta }E_{\rm x} W =  \dfrac{2  l_{\rm ee} E_{\rm x}W }{ \mathrm{sgn}(B) r_{\rm c}} = \dfrac{ 2 \tau_{ \rm ee} E_{\rm x} W  |{\rm e}| B }{ m_{\rm eff}}\\
& = \dfrac{ 12 \hbar^3 E_{\rm x} W  |{\rm e}|B }{ F_\pi^2 m_\mathrm{eff}^2 k_\mathrm{B}^2 T^2 }  \mathrm{ln}^2 \left( \frac{\hbar^2 \pi \rho_0}{m_\mathrm{eff} k_\mathrm{B} T} \right) \rho_0 = f_1 \rho_0 \ .  \nonumber 
\end{align}
Here, according to Refs.~\cite{Novikov06,PhysRevB.26.4421,alekseev2016negative,Qian05}, we introduced the second momentum relaxation time 
\begin{align}
\tau_{\mathrm{ee}}= \dfrac{l_\mathrm{ee}}{v_\mathrm{F}} = \dfrac{6 \hbar^3}{F_\pi^2 m_\mathrm{eff} k_\mathrm{B}^2 T^2} \mathrm{ln}^2 \left( \frac{\hbar^2 \pi \rho_0}{m_\mathrm{eff} k_\mathrm{B} T} \right) \rho_0 \ ,
\end{align}
where $k_\mathrm{B}$ is the Boltzmann constant and $F_\pi$ is a geometric factor, characterizing electron-electron scattering amplitudes. Since 
density variations do not significantly change the ln$^2(\mu/k_\mathrm{B}T)$ terms in the Fermi liquid regime $\mu \! \gg \! \mathrm{k}_\mathrm{B}T$, we treat $f_1$ as density independent function. 
Moreover, the impurity contribution to $\Delta V_{\eta_{\rm H}}$ is given by
\begin{align}
\! \Delta V_{\eta_{\rm H}}^\mathrm{imp} \! &= \!  - \dfrac{\rm \eta_{\rm H} }{ \eta }E_{\rm x} \dfrac{1   }{  12  l_\mathrm{G}^2 }
 \left( W^3 \! + \! 6 l_{\rm s} W^2 \! + \! 6 l_{\rm s}^2 W\right)  
 \\
\!  &= \! - 
 \dfrac{ \vert \mathrm{e}  \vert \! B m_\mathrm{eff}^2 \nu_0^2 n_\mathrm{imp}  E_{\rm x}  }{    3 \hbar^5  \pi    }
 \left( W^3 \! + \! 6 l_{\rm s} W^2 \! + \! 6 l_{\rm s}^2 W\right) \nonumber
 \! = \!  f_2 .
\end{align}
Here, we considered point like impurities with concentration $n_\mathrm{imp}$, scattering strength $\nu_0$ and inverse momentum relaxation  time $\tau_\mathrm{imp}^{_{-1}} \! = \! v_\mathrm{F} l_\mathrm{imp}^{_{-1}}\! = \! m_\mathrm{eff} \nu_0^2 \hbar^{-3} \rho_0 n_\mathrm{imp} $ \cite{bookref}.
In the same manner, Eq.~\eqref{eq:lorentz} evolves to
\begin{align}
\Delta 	V_{B} \! &= \!  - \frac{ \mathrm{sgn}(B)  E_\mathrm{x}}{3 r_\mathrm{c} l_\mathrm{ee} }  \left( W^3 \! + \! 6 l_{\rm s} W^2 \! + \! 6 l_{\rm s}^2 W\right) \\
 & = - \frac{  \vert {\rm e} \vert m_\mathrm{eff}  B  E_\mathrm{x}}{6 \pi \hbar^2 \rho_0  \tau_\mathrm{ee}} \left( W^3 \! + \! 6 l_{\rm s} W^2 \! + \! 6 l_{\rm s}^2 W\right)  \nonumber
 = f_3 \rho_0^{-2} \ .
\end{align}

\noindent
After having clarified how to disentangle $\Delta 	V_{\eta_{\rm H}}$ and $\Delta	V_{B}$ in terms of their width and density dependence, let us proceed in presenting  our solution for $v^1_{\rm x}(y)$. Therefore, we substitute Eq.~\eqref{eq:WeakB mu} into Eq.~\eqref{eq:WeakB NSE3}, which leads to the first order velocity correction 
\begin{align}
\label{eq:WeakB vx1} 
v_{\mathrm{x}}^1(y)  = & \, ( C_1  +   \lambda_1 \, y) \cosh \! \left[yl^{-1}_{\rm G}\right]  \\
+ & \, (C_2  +   \lambda_2 \, y) \, \sinh \! \left[y l^{-1}_{\rm G} \right] - A_3 + A_4 \, y  . \nonumber
\end{align}
Here, for clarity, we defined the following functions

\allowdisplaybreaks
\begin{widetext}
\begin{align}
\label{eq:TheLambdas}
\lambda_{1,2}  = & \dfrac{{\rm e}^2  l_{\rm imp} E^2_\mathrm{x}\frac{\partial {\rho}_0}{\partial \mu} }{ 2 m_{\rm eff} v_{\rm F}} \left(  {\rm e}B  - \dfrac{m_{\rm eff} v_{\rm F}\eta_\mathrm{H} }{ l_{\rm imp}}\right)A_{2,1} \ \, \ , \ \, \quad
A_3  =  \dfrac{\mathrm{e} \, l_{\rm imp} E_\mathrm{x} \frac{\partial {\rho}_0}{\partial \mu} }{  m_{\rm eff}v_{\rm F} \rho_0 }\Gamma \ \, \ , \ \, \quad A_4 = \dfrac{\mathrm{e}^2 l^2_{\rm imp}  E_{\rm x} B }{ m^2_{\rm eff} v^2_{\rm F}\rho_0 } \ ,  \\
\nonumber 
\\
\label{eq:vx1 constants}
C_1 =&\, \text{csch}[(2 l_\mathrm{s} \! + \! W)l_{\rm G}^{-1}] \Bigl(\sinh[(l_\mathrm{s} \! + \! W)l_{\rm G}^{-1}]\left[A_3 \! - \! l_\mathrm{s} \left(\lambda _2 \sinh [l_\mathrm{s} l_{\rm G}^{-1}] \! - \! \lambda _1\cosh [l_\mathrm{s}l_{\rm G}^{-1}] \! - \! A_4\right)\right] 
\\
&-\sinh [l_\mathrm{s}l_{\rm G}^{-1}] \bigl[(\lambda _2 l_\mathrm{s} \!  - \! A_3 \! + \! W  ) \sinh [(l_\mathrm{s} \! + \! W)l_{\rm G}^{-1}] \! + \! \lambda _1 l_\mathrm{s} \! + \! W) \cosh [(l_\mathrm{s} \! + \! W)l_{\rm G}^{-1}]\! + \! A_4 l_\mathrm{s} \! + \! W)\bigr]\Bigr), \nonumber
\\
\nonumber
\\ 
C_2 =&  - \text{csch}[(2 l_\mathrm{s} \! + \! W)l_{\rm G}^{-1}] \Bigl(2 ( A_4 W  \! - \! A_3) \cosh [l_\mathrm{s} l_{\rm G}^{-1}] \! + \! 2 A_3 \cosh[(l_\mathrm{s} \! + \! W)l_{\rm G}^{-1}]
\\ &
+\lambda _2 W \sinh [(2 l_\mathrm{s} \! + \! W)l_{\rm G}^{-1}] 
\! + \! \sinh [Wl_{\rm G}^{-1}]
\!  + \! \lambda _1 W \bigl[\cosh [(2 l_\mathrm{s} \! + \! W)l_{\rm G}^{-1}] \! + \! \cosh [Wl_{\rm G}^{-1}]\bigr]\nonumber \\
 \! &+\! 2 l_\mathrm{s} \Bigl[\lambda _1 \bigl[\cosh [(2 l_\mathrm{s} \! + \! W)l_{\rm G}^{-1}]+\cosh [Wl_{\rm G}^{-1}]\bigr] \!+ \! A_4 \bigl[\cosh [(l_\mathrm{s} \! + \! W)l_{\rm G}^{-1}]\! + \! \cosh [l_\mathrm{s} l_{\rm G}^{-1}]\bigr]\! + \! \lambda _2 \sinh [Wl_{\rm G}^{-1}]\Bigr]\Bigr)/2 \ , \nonumber
\end{align}
\end{widetext}
Notice, that for $\eta_{\rm H} \neq 0$, the first order correction $v_{\mathrm{x}}^1(y)$ breaks the reflection symmetry of the entire velocity profile with respect to $y=W/2$, as expected. Explicitly this is caused by the linear terms in powers of $y$ within Eq.~\eqref{eq:WeakB vx1}. Moreover, since  $C_{1,2}, \, \lambda_{1,2}$ and $A_{1,2,3,4}$ are  non-linear in $E_{\rm x}$ and $l_{\rm imp}$, Eq.~\eqref{eq:WeakB vx1} implies that the first order correction $v^1_{\rm x} \ll v^0_{\rm x}$ in the linear response regime.


\textit{Strong magnetic field regime.} 
We now move on to the discussion of strong magnetic fields, implicitly defined by $r_\mathrm{c} \ll l_\mathrm{G}$.
In this limit,  $\eta$ and $\eta_\mathrm{H}$ tend to zero (cf. Eq.~(4) of our main text), which strongly simplifies \mbox{Eqs.~\eqref{eq:NPNS1} and \eqref{eq:NPNS2}}. In particular, these equations yield 
\begin{align}
\label{eq:SBNS1}
v_{\rm x}=-\dfrac{{\rm e} \, l_{\rm imp}E_{\rm x} }{  m_{\rm eff} v_{\rm F}} \quad \wedge \quad \partial_\mathrm{y} \Delta \mu = {\rm e}Bv_{\rm x} \ ,
\end{align}
Here, we dropped the term $\propto \! \! \Delta \mu \, E_{\rm x}$ to make explicit predictions for our linear response theory. In particular, the solution of Eq.~\eqref{eq:SBNS1} reads:
\begin{align} \label{eq:Strong B DV} 
\Delta  V_\mathrm{tot} =  -\mathrm{sgn}(B)  E_{\rm x}  W l_{\rm imp}  / r_\mathrm{c} \ .
\end{align}
Hence, the Hall voltage solely depends on $l_\mathrm{imp}$ and $W$ in the limit of strong magnetic fields.

\textit{Robin Boundary Conditions.}
In what follows, we prove that our set of boundary conditions in Eq.~\eqref{eq:bcs} is equivalent to Robin type  boundary conditions, satisfying
\begin{align}
\label{eq:RobinBCs} u^0_{\rm x}(0) - l_R \partial_y u_{\rm x}(0) =  u_{\rm x}(W) + l_R\partial_y u_{\rm x}(W) = 0 \ .
\end{align} 
Here $u^0_{\rm x}(y)$ defines the Robin velocity profile  and  $l_R$ is the corresponding slip length. As a solution of the Navier-Stokes Eqs.~\eqref{eq:WeakB NSE1}-\eqref{eq:WeakB NSE3}, $u^0_{\rm x}(y)$  is given by
\begin{align}
\label{eq:WeakB ux0}
u^0_{\rm x}(y) \! = \! - \dfrac{{\rm e} E_{\rm x}l_{\rm imp} }{ m_{\rm eff}v_{\rm F}} \! \! \left(  C_1  \cosh \! \left[   yl_{\rm G}^{-1}  \right] \!  +  C_2  \sinh \! \left[   yl_{\rm G}^{-1}  \right] \! \! + \! 1   \right)  \! . 
\end{align}
The amplitudes $C_{1}$ and  $C_{2}$ are defined to be:
\begin{align}
\label{eq:ux0 constants} 
C_1 & = -l_{\rm G}\left[l_{\rm G} + l_R \tanh\left[W /( 2l_{\rm G})\right]\right]^{-1} \ , 
\\
C_2 & = \ \; \;  l_{\rm G}\left[l_R + l_{\rm G}{\rm coth}\left[W /( 2l_{\rm G}) \right]\right]^{-1} \ \, .
\end{align} 
In order to prove the equivalence between the two types of boundary conditions, it is sufficient to show that the velocity profiles $u^0_{\rm x}(y)$ and  $v^0_{\rm x}(y)$ match each other.  According to  Eq.~\eqref{eq:WeakB ux0}, the Robin slip length is defined to be the normalized value of the inverse first derivative of the velocity profile at the boundaries. Hence, let us define the effective Robin slip length for our velocity profile
\begin{align} \label{deflbeff}
l_R^\mathrm{eff} \equiv {v^0(0) \over \partial_y v^0_{\rm x}(0)} = - {v^0(W) \over \partial_y v^0_{\rm x}(W)} \ .
\end{align}
This quantity allows us to derive the equivalence of the two types of boundary conditions if we can prove that
\begin{align} \label{mappingvelocity}
u^0_{\rm x}(y; l_R=l_R^\mathrm{eff}) = v^0_{\rm x}(y;l_{\rm s}) \ .
\end{align} 
For $l_\mathrm{G}/r_\mathrm{c} \ll 1$, Eq.~\eqref{eq:WeakB vx0}
defines  $v^0_{\rm x}(y)$ analytically. In this limit, Eq.~\eqref{deflbeff} implies for $l_R=l_R^\mathrm{eff}$:
\begin{align} \label{eq:Effective lR} 
l_R  & =l_{\rm G} \ {\rm csch}[W \! / l_{\rm G}] \! \left(\cosh[(2l_s \!  + \!  W)/l_{\rm G}] \! - \! \cosh[W/(2l_{\rm G})]\right) \! ,\nonumber \\
l_{\rm s} &= l_{\rm G} \mathrm{arcosh}  [ \sinh [ W/(2l_{\rm G})] \left(l_{\rm G} \! \coth [  W/(2l_{\rm G}) ]  \! + \! l_R \right)\! /l_{\rm G}] \nonumber 
\\ & \quad - W/2  \ .
\end{align}
In fact, this mapping identifies the velocity profiles
$u^0_{\rm x}(y)$ and  $v^0_{\rm x}(y)$, since it maps  Eq.~\eqref{eq:WeakB ux0} onto  Eq.~\eqref{eq:WeakB vx0}. 
In particular, the Robin type equivalents to the Hall viscous and the Lorentz force contribution to the total Hall voltage  in Eqs.~\eqref{eq:HallVoltagesCase32} and \eqref{eq:lorentz} are given by
\begin{align}
\label{eq:Robin voltages}
\Delta 	V_{\eta_{\rm H}}^R &=  \frac{\eta_\mathrm{H} }{\eta}  E_\mathrm{x} \left[ W -  
  { 1  \over  12  l_\mathrm{G}^2 }
 \left( W^3 \! + \! 6 l_R W^2 \right) \right] ,\nonumber \\
\Delta 	V_{B}^R &=  - \frac{ \mathrm{sgn}(B)  E_\mathrm{x}}{3 r_\mathrm{c} l_\mathrm{ee} }  \left( W^3 \! + \! 6 l_R W^2\right) \ .
\end{align}
Since Eq.~\eqref{eq:HallVoltagesCase32} and 
Eq.\eqref{eq:lorentz} are derived under the assumption $W / l_{\rm G} \ll 1$, Eq.\eqref{eq:Robin voltages} is obtained by inserting 
the Robin slip length, given by Eq. \eqref{eq:Effective lR} in the same limit, $l_R \simeq l_{\rm s} + l^2_{\rm s}/W$.
Beyond the weak magnetic field regime Eq.~\eqref{mappingvelocity} is still justified. 
Since in this limit $l_\mathrm{G} / r_\mathrm{c} \gtrsim 1$, this can not be proven analytically. Instead, we confirmed the validity of Eq.~\eqref{mappingvelocity} by numerically evaluating the Robin slip length in Eq.~\eqref{deflbeff}. 
As soon as the cyclotron radius $r_\mathrm{c} \ll W$, one enters the strong magnetic field regime. While in this limit the Robin slip length diverges, our slip length remains constant by definition. Therefore, we can not justify Eq.~\eqref{mappingvelocity} for strong magnetic fields. However, in this limit the equivalence of both types of boundary conditions is given trivially. Since the fluid layer interaction decreases as function of $r_\mathrm{c}$, all our results become slip length and therefore boundary condition independent for $r_\mathrm{c} \ll W$ (cf.~Eq.~\eqref{eq:Strong B DV}). 

\end{document}